\documentclass[usenatbib]{mnras}
\usepackage{aas_macros}
\usepackage{amsfonts}
\usepackage{amsmath}
\usepackage{amssymb}
\usepackage{graphicx}
\usepackage[dvipsnames]{xcolor}
\hypersetup{colorlinks=true,allcolors=teal}

\newcommand{\NSIM}{\ensuremath{10}}

\newcommand{\p}{\ensuremath{\partial}}

\newcommand{\Mh}{\ensuremath{h^{-1}M_{\odot}}}

\newcommand{\Mpch}{\ensuremath{h^{-1}{\rm Mpc}}}
\newcommand{\kpch}{\ensuremath{h^{-1}{\rm kpc}}}

\newcommand{\avg}[1]{\ensuremath{\left\langle \,#1\, \right\rangle}}
\newcommand{\e}[1]{\ensuremath{{\rm e}^{#1}}}

\newcommand{\der}{\ensuremath{{\rm d}}}
\newcommand{\dir}{\ensuremath{\delta_{\rm D}}}

\newcommand{\dH}{\ensuremath{\delta_{\rm halo}}}

\newcommand{\eqn}[1]{equation~\eqref{#1}}

\newcommand{\ph}[1]{\phantom{#1}}

\newcommand{\be}{\begin{equation}}
\newcommand{\ee}{\end{equation}}
\newcommand{\Cal}[1]{\ensuremath{\mathcal{#1}}}

\title[Assembly bias and tidal anisotropy]
      {Halo assembly bias and the tidal anisotropy of the local halo environment} 
\date{draft}

\author[Paranjape, Hahn \& Sheth]{
Aseem Paranjape$^{1}$\thanks{E-mail: aseem@iucaa.in}, 
Oliver Hahn$^{2}$\thanks{E-mail: oliver.hahn@oca.eu} 
\& Ravi K. Sheth$^{3,4}$\thanks{E-mail: shethrk@physics.upenn.edu}
\\  
 $^1$ Inter-University Centre for Astronomy \& Astrophysics,
      Ganeshkhind, Post Bag 4, Pune 411007, India\\
 $^2$ Laboratoire Lagrange, Universit\'e C\^ote d'Azur, Observatoire de la C\^ote d'Azur, CNRS, 
      Blvd de l'Observatoire,\\\hskip0.15in CS 34229, 06304 Nice cedex 4, France\\
 $^3$ Center for Particle Cosmology, University of Pennsylvania, 
      209 S. 33rd St., Philadelphia, PA 19104, USA\\
 $^4$ The Abdus Salam International Center for Theoretical Physics,
      Strada Costiera, 11, Trieste 34151, Italy}

\begin{document}

\label{firstpage}
\pagerange{\pageref{firstpage}--\pageref{lastpage}}

\maketitle 

\begin{abstract}
\noindent
We study the role of the local tidal environment in determining the assembly bias of dark matter haloes. 
Previous results suggest that the anisotropy of a halo's environment (i.e, whether it lies in a filament or in a more isotropic region) can play a significant role in determining the eventual mass and age of the halo. 
We statistically isolate this effect using correlations between the large-scale and small-scale environments of simulated haloes at $z=0$ with masses between $10^{11.6}\lesssim (m/h^{-1}M_{\odot})\lesssim10^{14.9}$. 
We probe the large-scale environment using a novel halo-by-halo estimator of linear bias. 
For the small-scale environment, we identify a variable $\alpha_R$ that captures the \emph{tidal anisotropy} in a region of radius $R=4R_{\rm 200b}$ around the halo and correlates strongly with halo bias at fixed mass. 
Segregating haloes by $\alpha_R$ reveals two distinct populations. 
Haloes in highly isotropic local environments ($\alpha_R\lesssim0.2$) behave as expected from the simplest, spherically averaged analytical models of structure formation, showing a \emph{negative} correlation between their concentration and large-scale bias at \emph{all} masses. 
In contrast, haloes in anisotropic, filament-like environments ($\alpha_R\gtrsim0.5$) tend to show a \emph{positive} correlation between bias and concentration at any mass. 
Our multi-scale analysis cleanly demonstrates how the overall assembly bias trend across halo mass emerges as an average over these different halo populations, and provides valuable insights towards building analytical models that correctly incorporate assembly bias. 
We also discuss potential implications for the nature and detectability of galaxy assembly bias. 
\end{abstract}

\begin{keywords}
cosmology: theory, dark matter, large-scale structure of the Universe -- methods: numerical
\end{keywords}

\section{Introduction}
\label{sec:intro}
\noindent
The assembly history of dark matter haloes is known to correlate with the large scale environment, even for haloes of fixed current mass \citep{st04,gsw05}. This effect, known as \emph{halo assembly bias}, has been well studied in the literature using $N$-body simulations, and extends to several halo properties including age, accretion rate, concentration, spin, shape, velocity dispersion and anisotropy, etc. \citep[see, e.g.][]{wechsler+06,jsm07,desjacques08,hahn+09,fm10,fw10,swm15,bprg17,pp17,lms17}. 
To the extent that galaxy formation and evolution is regulated by the accretion of dark matter onto the host halo of a galaxy, assembly bias can in principle have interesting observational consequences \citep{zhv14,hbv16}. While there have been several recent observational attempts at detecting assembly bias effects in galaxy and cluster populations \citep{lin+16,miyatake+16,montero-dorta+17}, systematic effects in cleanly segregating galaxy populations have been challenging to overcome \citep{twcm17,zu+17}. Recent high-resolution hydrodynamical simulations of galaxy assembly based on small galaxy samples seem to be consistent with small/negligible galaxy assembly bias effects \citep{rgbp17,grbp17}, while hydrodynamical simulations of cosmological volumes have led to results qualitatively similar to their dark matter only counterparts \citep[see, e.g.,][]{chaves-montero+16,bray+16}.

The assembly bias trend seen in numerical studies is that, at large halo mass, highly concentrated or old haloes cluster weakly as compared to less concentrated or younger haloes of the same mass. At low mass on the other hand, the trend inverts, with old haloes clustering \emph{more} strongly than younger ones. 
The trend for massive haloes is, in fact, qualitatively predicted by simple models of structure formation based on peaks theory \citep{dwbs08}, ellipsoidal dynamics \citep{desjacques08} or the excursion set formalism \citep{ms12,cs13}. 
In these models, assembly bias arises from a strong correlation between the density structure of a Lagrangian `proto-halo' patch that is destined to become a virialised halo and its larger scale density environment. These strong correlations naturally produce haloes with large inner density (or high concentration) which form early and live in more underdense environments as compared to haloes of the same mass but with lower inner density, that form late and live in denser environments. The peaks and/or excursion set models, however, predict that this qualitative trend should be seen at \emph{any} fixed mass, and therefore contradict the inversion of the trend around $m\sim m_\ast$ as seen in simulations, where $m_\ast$ is the scale where the mass fraction in haloes has a maximum. The inversion is therefore likely to be associated with some key physical criterion that the models are curently missing.

One plausible mechanism for explaining the inversion was proposed by \citet{hahn+09}. These authors argued that low mass haloes in filaments suffer tidal truncation of their mass accretion due to redirected mass flows along their filament, with the degree of truncation depending on the magnitude of anisotropic velocity shear in the vicinity of the halo. This has been further established by \citet{bprg17} using zoom simulations of individual low mass haloes in a variety of tidal environments \citep[see also][]{behroozi+14}. In the present work, we further explore the role of the tidal environment in establishing the inversion of the low mass assembly bias trend using a multi-scale statistical study of $N$-body simulations. Our aim is to identify the key statistical variable (there may be more than one) that controls the sign of the correlation between halo properties such as age or concentration and their large scale bias. Our expectation is that such a study will provide clear guidance for how tidal effects might be included in analytical models of halo abundances and clustering that correctly incorporate assembly bias at all mass scales.

The paper is organised as follows. In section~\ref{sec:numerics} we describe our $N$-body simulations and the construction of an object-by-object linear bias estimator $b_1$ that becomes a valuable tool for studying multi-scale correlations. In section~\ref{sec:bias-env} we establish a statistical link between the large scale halo environment (as measured by $b_1$) and the \emph{anisotropy} of the halo's local tidal environment. In particular, we identify a scalar combination $\alpha_R$ of the eigenvalues of the tidal tensor smoothed on scale $R=4R_{\rm 200b}$\footnote{$R_{\rm 200b}$ is defined as the radius where the enclosed density is $200$ times the background density. The mass enclosed inside $R_{\rm 200b}$ is denoted $m_{\rm 200b}$.} at the halo location and show that $\alpha_R$ correlates more strongly with $b_1$ than does the overdensity $\delta_R$ on the same scale. Then, in section~\ref{sec:assemblybias}, we study halo assembly bias as a function of tidal anisotropy, by measuring the correlation between $b_1$ and halo concentration (a proxy for halo age) as a function of $\alpha_R$. This allows us to link halo internal properties with both the small scale tidal environment and the large scale density around the halo. We discuss the implications of this multi-scale study for understanding the origin of assembly bias in section~\ref{sec:discuss}, and conclude in section~\ref{sec:conclude}. The Appendices give technical details of some of the results used in the main text.

Throughout, we use a spatially flat Lambda cold dark matter ($\Lambda$CDM) cosmology with total matter density parameter $\Omega_{\rm m}=0.276$, baryonic matter density $\Omega_{\rm b}=0.045$, Hubble constant $H_0=100h\,{\rm kms}^{-1}{\rm Mpc}^{-1}$ with $h=0.7$, primordial scalar spectral index $n_{\rm s}=0.961$ and r.m.s. linear fluctuations in spheres of radius $8\Mpch$, $\sigma_8=0.811$, with a transfer function generated by the code \textsc{camb} \citep{camb}.\footnote{http://camb.info}

\section{Numerical techniques}
\label{sec:numerics}
\noindent
Below, we describe the $N$-body simulations used in this work, followed by a description of a novel object-by-object estimator of halo clustering that we will use in our analysis.

\subsection{$N$-body simulations}
\label{subsec:sims}
\noindent
We have performed $N$-body simulations of CDM using the tree-PM code \textsc{gadget-2} \citep{springel:2005}\footnote{http://www.mpa-garching.mpg.de/gadget/} with $N_{\rm p}=1024^3$ particles in a cubic, periodic box. 
We use two configurations: a lower resolution one for which we generate \NSIM\ realisations, and a single realisation of a smaller volume, higher resolution box. The details of these configurations are given below.

Our lower resolution configuration uses a box of comoving length $L_{\rm box}=300\Mpch$ and a $2048^3$ PM grid, with force resolution $\epsilon=9.8\,\kpch$ comoving. For our chosen cosmology, this gives a particle mass of $m_{\rm p}=1.93\times10^9\Mh$. As we will see, this configuration allows us to straddle the characteristic mass scale $m_\ast$ of the halo mass function at $z=0$ with sufficient dynamic range to probe both the regimes of assembly bias mentioned earlier. Initial conditions were generated at a starting redshift $z_{\rm in}=49$ using the code \textsc{music} \citep{hahn11-music}\footnote{https://www-n.oca.eu/ohahn/MUSIC/} with 2nd order Lagrangian perturbation theory (2LPT). Haloes were identified using the code \textsc{rockstar} \citep{behroozi13-rockstar}\footnote{http://code.google.com/p/rockstar/} which performs a Friends-of-Friends (FoF) algorithm in 6-dimensional phase space. The simulations and analysis were performed on the Perseus cluster at IUCAA.\footnote{http://hpc.iucaa.in}

To ensure that our results are not contaminated by substructure and numerical artefacts, we discard all sub-haloes and further only consider objects whose `virial' energy ratio $\eta=2T/|U|$ satisfies $0.5\leq\eta\leq1.5$ as suggested by \citet{Bett+07}. Below, we will heavily rely on measurements of the tidal environment in the vicinity of the haloes. These measurements are performed after Gaussian smoothing on a cubic grid with $N_{\rm g}=512^3$ cells. We consider multiple choices of smoothing scales as described later in the text. We therefore impose a restriction on the minimum halo mass we study, so as to minimise the contamination to our final results from the resolution imposed by this grid. We describe our procedure in Appendix~\ref{app:scaling}; this leads to a minimum halo mass $m_{\rm 200b} \geq m_{\rm min} \simeq 3.1\times10^{12}\Mh$, corresponding to haloes resolved with $N_{\rm p}^{\rm (halo)}\geq1600$ particles each. These cuts leave us with approximately $38,700$ objects on average at $z=0$ in a single realisation of the simulation. Additionally, throughout the analysis we impose an upper limit of $m_{\rm 200b}\leq m_{\rm max}=7.7\times10^{14}\Mh$, corresponding to the mass scale above which we expect fewer than $10$ haloes for our box size and cosmology at $z=0$. To improve our statistics, we have generated \NSIM\ realisations of our simulation by changing the seed for the initial conditions. 

We will also additionally use the output of a single realisation of a simulation with the same cosmology, number of particles and PM grid, but having $L_{\rm box}=150\Mpch$ and a force resolution $\epsilon=4.9\,\kpch$, which will extend our mass range down to $m_{\rm 200b} \gtrsim 3.85\times10^{11}\Mh$. We will refer to this as the high resolution box, and to the \NSIM\ larger volume realisations as the default box. We will use comparisons between the statistics inferred from these two boxes to demonstrate the numerical convergence of our results. Throughout, we will focus on results at $z=0$.

\subsection{Halo-by-halo estimator of bias}
\label{subsec:HbHbias}
\noindent
Traditional estimators of halo bias involve ratios of (cross) power spectra of haloes and dark matter. Exploiting some basic properties of discrete Fourier transforms, we construct an object-by-object estimator of large-scale linear halo bias, whose average properties reproduce known trends derived from traditional estimators. This new halo-by-halo bias then becomes a useful probe of the correlations between large-scale and small-scale halo environment, and between these two and other halo properties such as assembly history, halo (sub-)structure, etc. We give the details of our construction below.

The traditional cross-correlation based estimator of halo bias in Fourier space is the ratio of the halo-matter cross power spectrum $P_{\rm hm}(k)$ and the matter auto power spectrum $P_{\rm mm}(k)$:
\be
b_{\rm hm}(k) \equiv P_{\rm hm}(k)/P_{\rm mm}(k)
\label{eq:b1trad}
\ee
At large scales ($k\to0$), this recovers the `peak-background split' value of Eulerian linear bias \citep{ps12a,sjd13}.
It is instructive to recapitulate the procedure for deriving the halo-matter cross power spectrum in a simulation box. In the following, we will consider a collection of haloes indexed by the integer variable $h$ whose values are restricted according to some condition \Cal{C}. E.g., \Cal{C} could refer to selecting haloes in a chosen mass bin.  Starting with the positions $\{\mathbf{x}_h\}$ of all haloes in a (cubic, periodic) simulation of volume $V_{\rm box}$ and a grid with $N_{\rm g}$ cubic cells, we define the number overdensity of \Cal{C}-haloes $\dH(\mathbf{x}|\Cal{C})$ at the grid cell with position $\mathbf{x}$ as
\begin{align}
\dH(\mathbf{x}|\Cal{C}) &\equiv n_{\rm halo}(\mathbf{x}|\Cal{C})/\bar n_{\rm halo}(\Cal{C}) - 1 \notag\\ 
&= \sum_{h\,\in\,\Cal{C}}\,N_{\rm g}\,\vartheta(\mathbf{x},\mathbf{x}_h)\,/\sum_{h\,\in\,\Cal{C}} \,-\, 1
\label{eq:dHgrid}
\end{align}
where $\vartheta(\mathbf{x},\mathbf{x}_h)$ is a selection function that gives the contribution of halo $h$ with position $\mathbf{x}_h$ to the cell at $\mathbf{x}$ and satisfies $\sum_{\{\mathbf{x}\}}\,\vartheta(\mathbf{x},\mathbf{x}_h)=1$ when summed over the grid, so that the \Cal{C}-halo number density is the sum over \Cal{C}-haloes\footnote{This is the discretized version of the continuum result $n_{\rm halo}(\mathbf{x}|\Cal{C})=\sum_{h\,\in\,\Cal{C}}\,\dir(\mathbf{x}-\mathbf{x}_h)$.} $n_{\rm halo}(\mathbf{x}|\Cal{C})=\sum_{h\,\in\,\Cal{C}}\,\vartheta(\mathbf{x},\mathbf{x}_h)$, with mean number density
\be
\bar n_{\rm halo}(\Cal{C}) = \sum_{\{\mathbf{x}\}} \,n_{\rm halo}(\mathbf{x}|\Cal{C}) \,/ \sum_{\{\mathbf{x}\}} = \sum_{h\,\in\,\Cal{C}}\,1/N_{\rm g}\,,
\label{eq:mean-nh}
\ee
since $\sum_{\{\mathbf{x}\}}\, = N_{\rm g}$.

The discrete Fourier transform of $\dH(\mathbf{x}|\Cal{C})$ can then be manipulated as follows:
\begin{align}
\dH(\mathbf{k}|\Cal{C}) 
&\equiv \frac{1}{N_{\rm g}}\,\sum_{\{\mathbf{x}\}}\,\e{i\mathbf{k}\cdot\mathbf{x}}\,\dH(\mathbf{x}|\Cal{C}) \notag\\
&= \left[\sum_{h\,\in\,\Cal{C}}\,\sum_{\{\mathbf{x}\}}\,\e{i\mathbf{k}\cdot\mathbf{x}}\,\vartheta(\mathbf{x},\mathbf{x}_h)\,/\sum_{h\,\in\,\Cal{C}}\right] - \sum_{\{\mathbf{x}\}}\,\e{i\mathbf{k}\cdot\mathbf{x}}/N_{\rm g}\notag\\
&= \sum_{h\,\in\,\Cal{C}}\,\e{i\mathbf{k}\cdot\mathbf{x}(h)} / \sum_{h\,\in\,\Cal{C}} \,-\, \delta^{\rm Kronecker}_{\mathbf{k},\mathbf{0}}\,,
\label{eq:dHFourier-calc}
\end{align}
where we used the shorthand notation $\e{i\mathbf{k}\cdot\mathbf{x}(h)}$ to denote the appropriate weighted sum of phase factors over all cells receiving a contribution from halo $h$.\footnote{Our notation corresponds to the exact result for the nearest grid point (NGP) scheme, with $\mathbf{x}(h)$ in this case being the location of the single cell that contains halo $h$. For the cloud-in-cell (CIC) scheme, which we use in practice, $\e{i\mathbf{k}\cdot\mathbf{x}(h)}$ stands for a weighted sum over eight cells.} 
Ignoring the Kronecker delta which enforces $\dH(\mathbf{k}=\mathbf{0}|\Cal{C})=0$, we are left with
\be
\dH(\mathbf{k}|\Cal{C}) = \sum_{h\,\in\,\Cal{C}}\,\e{i\mathbf{k}\cdot\mathbf{x}(h)} / \sum_{h\,\in\,\Cal{C}} \,\textrm{,~~~ for } \mathbf{k}\neq\mathbf{0}\,.
\label{eq:dHFourier}
\ee
A similar calculation holds for the matter density fluctuation field $\delta(\mathbf{k})$, but we will not need its explicit form below.

The required power spectra then follow from taking averages in spherical shells of $\mathbf{k}$; denoting these by $\avg{}_k$, we have
\begin{align}
P_{\rm hm}(k|\Cal{C}) &= V_{\rm box}\,\avg{\dH(\mathbf{k}|\Cal{C})\delta^\ast(\mathbf{k})}_k \notag\\
&= V_{\rm box}\,\sum_{h\,\in\,\Cal{C}}\avg{\e{i\mathbf{k}\cdot\mathbf{x}(h)}\delta^\ast(\mathbf{k})}_k/\sum_{h\,\in\,\Cal{C}} \,,\notag\\
P_{\rm mm}(k) &= V_{\rm box}\,\avg{\delta(\mathbf{k})\delta^\ast(\mathbf{k})}_k \,,
\label{eq:Pk-def}
\end{align}
where the asterisk denotes a complex conjugate.
The expression \eqref{eq:b1trad} for $k$-dependent linear bias of \Cal{C}-haloes then reduces to
\begin{align}
b_{\rm hm}(k|\Cal{C}) &= \sum_{h\,\in\,\Cal{C}} \left(\frac{V_{\rm box}}{P_{\rm mm}(k)}\, \avg{\e{i\mathbf{k}\cdot\mathbf{x}(h)}\delta^\ast(\mathbf{k})}_k\right)/\sum_{h\,\in\,\Cal{C}} \notag\\
&\equiv \sum_{h\,\in\,\Cal{C}}\,b_{1,h}(k) / \sum_{h\,\in\,\Cal{C}}
\label{eq:b1k}
\end{align}
where the second line defines an object-by-object, scale dependent quantity $b_{1,h}(k)$ whose average over the haloes under consideration corresponds to the usual scale-dependent cross-correlation linear bias. Notice that the selection criterion \Cal{C} only appears in defining the average by restricting the summation range.

We can reduce $b_{1,h}(k)$ to a single number for each halo by averaging over low-$k$ modes\footnote{For the analysis in this paper, we use $0.025\lesssim k/(h{\rm Mpc}^{-1})\lesssim0.09$ for our default box, and $0.05\lesssim k/(h{\rm Mpc}^{-1})\lesssim0.09$ for the high resolution box.} for which the bias is expected to be nearly constant: 
\begin{align}
b_{1,h} &\equiv \sum_{\textrm{low }k}\,N_k\,b_{1,h}(k) / \sum_{\textrm{low }k}\,N_k \notag\\
&= \sum_{\textrm{low }k} \,N_k\left(V_{\rm box}\, \avg{\e{i\mathbf{k}\cdot\mathbf{x}(h)}\delta^\ast(\mathbf{k})}_k/P_{\rm mm}(k)\right) / \sum_{\textrm{low }k}\,N_k\,,
\label{eq:b1HbyH}
\end{align}
where we have weighted by the number of modes $N_k\propto k^3$ for logarithmically spaced bins.
We will refer to this quantity $b_{1,h}$, defined for every halo $h$, as halo-by-halo bias. For ease of notation, we will drop the subscript $h$ whenever no confusion can arise.

\begin{figure}
\centering
\includegraphics[width=0.45\textwidth]{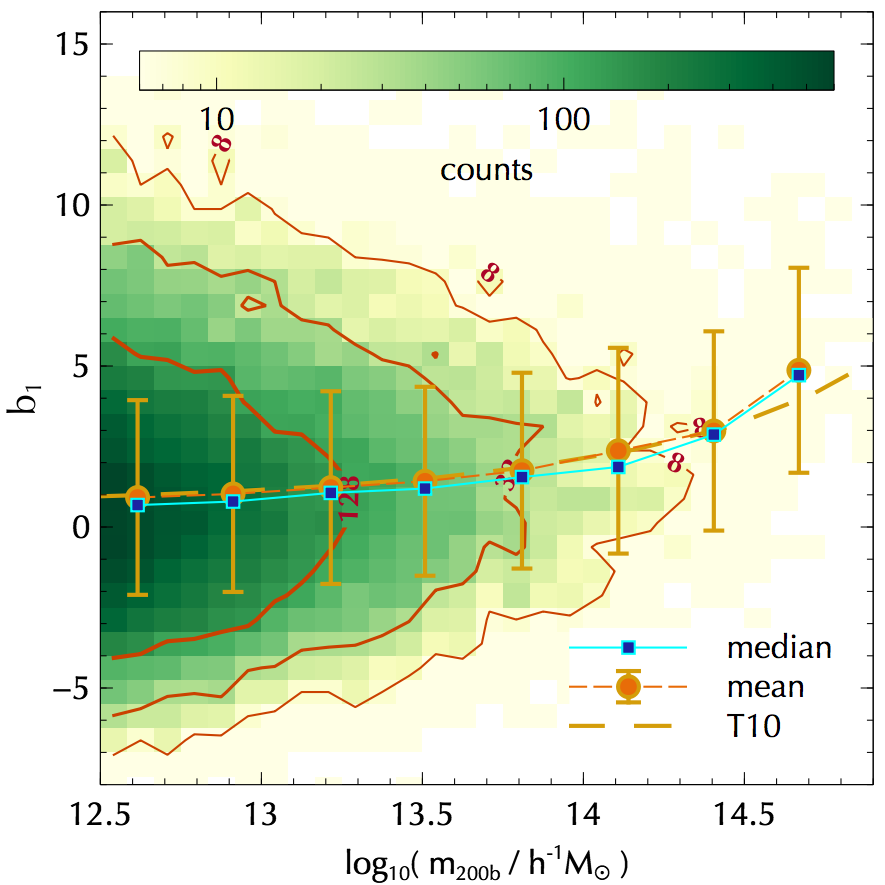}
\caption{Halo-by-halo bias. Coloured region shows the distribution $b_1$ as a function of halo mass $m=m_{\rm 200b}$ for haloes in one realisation of our default box, with $b_1$ evaluated for individual haloes using \eqn{eq:b1HbyH}. The colour indicates the number of haloes for this realisation in each 2-dimensional bin and the contours indicate bins of fixed number counts (8, 32, 128) as labelled. Points show the median (filled blue squares) and mean (filled orange circles) of $b_1$ in bins of halo mass. Error bars indicate the scatter (standard deviation) in each mass bin. Smooth dashed curve shows the fitting function for linear bias appropriate for $m_{\rm 200b}$-haloes taken from \citet[][T10]{Tinker10}.}
\label{fig:b1Vsmass-basic}
\end{figure}

This definition of halo-by-halo bias has several useful properties. Firstly, our derivation above shows that $b_1$ averages to the usual peak-background split bias for any choice of halo selection criterion \Cal{C} (e.g., binning by mass). Being defined for each halo, however, makes $b_1$ a convenient additional property that can be included in a halo catalog and studied in conjunction with any other halo property of interest, \emph{without} any need for binning in principle. 
The coloured region and contours in Figure~\ref{fig:b1Vsmass-basic} show the distribution of halo mass and halo-by-halo bias computed for individual haloes in one simulation box. 
The filled blue and orange points respectively show the median and mean of $b_1$ in bins of halo mass, while the dashed orange curve shows the fitting function from \citet{Tinker10}.
We see that, as expected, there is good agreement between the measurements and the fit at all but the highest masses which suffer from volume effects and possibly also some mild scale dependence due to our choice of $k$ range. 

\begin{figure}
\centering
\includegraphics[width=0.45\textwidth]{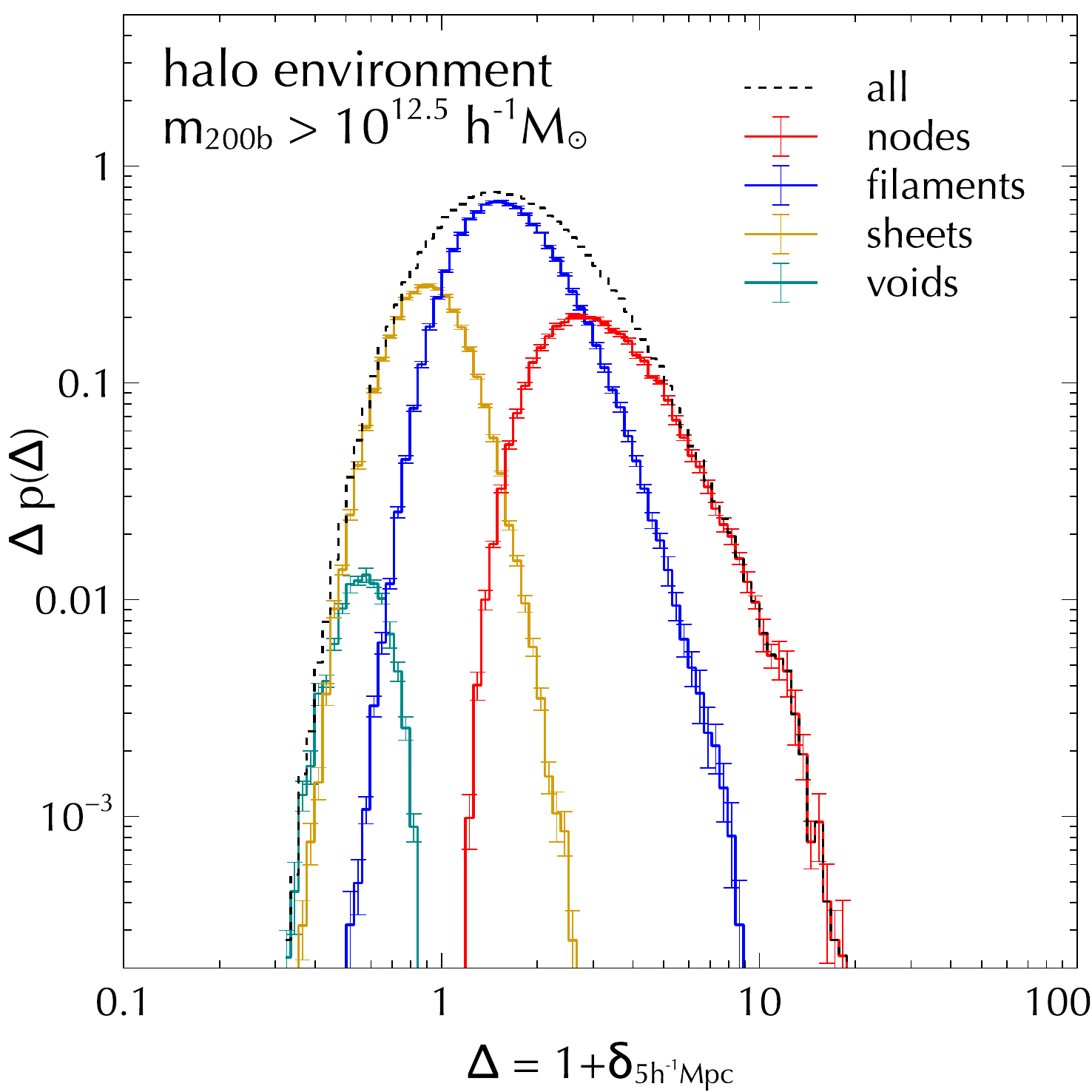}
\caption{Dark matter density in halo environments. Histograms show the distribution of overdensity $1+\delta_R$ centered on haloes selected as described in the text and smoothed with a Gaussian window of radius $R=5\Mpch$, averaged over \NSIM\ realisations of the default box. The error bars indicate the standard error on the mean over the \NSIM\ realisations. Dashed histogram shows the total distribution, while the various colours indicate different halo environments (from right to left in order of peak location: nodes, filaments, sheets and voids).}
\label{fig:hist-delta5Mpch}
\end{figure}

The most striking aspect of the Figure, however, is the large scatter in $b_1$. The standard deviation of $b_1$, shown by the error bars on the measurements of the mean, is about $\sigma_{b_1}\simeq3$ at essentially all masses. (Note that the error on the mean is much smaller, due to the large number of points in each bin.) At low masses, this means that the small value of mean or median halo bias is, in fact, a rather poor indicator of the large scale environment of these objects. We give an analytical argument explaining this large value of the scatter in Appendix~\ref{subapp:b1-delta}. Below, we will explore the relation between this scatter in $b_1$ and the properties of the local environment of the haloes populating the tails of the distribution.

\begin{figure*}
\centering
\includegraphics[width=0.45\textwidth]{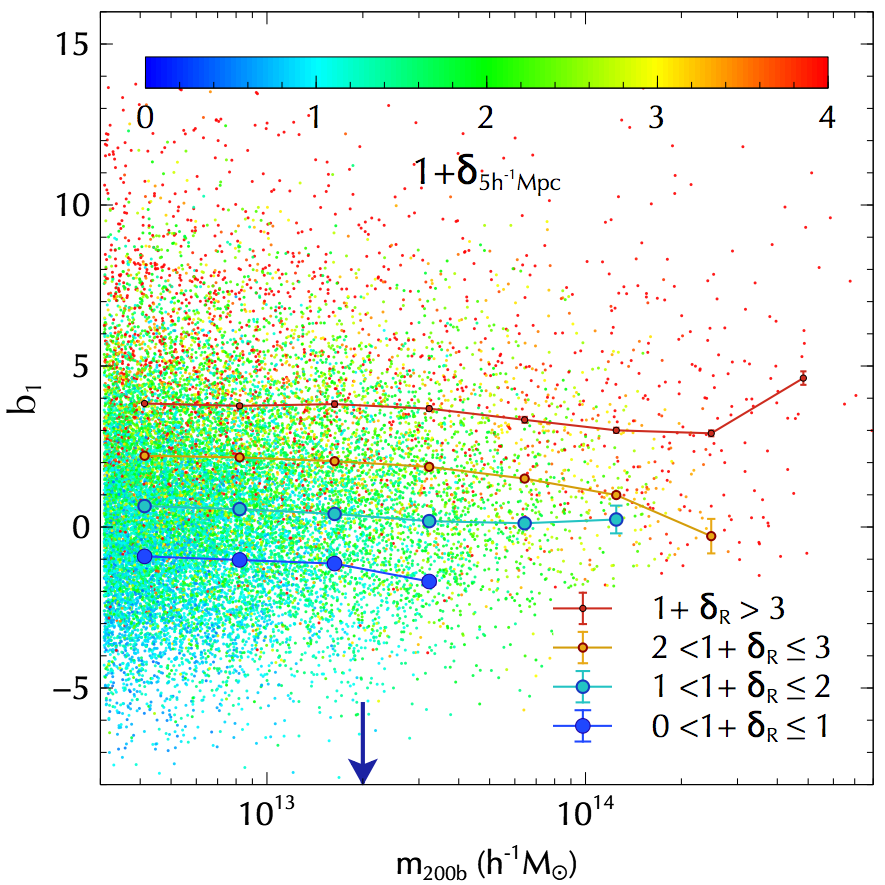}
\includegraphics[width=0.45\textwidth]{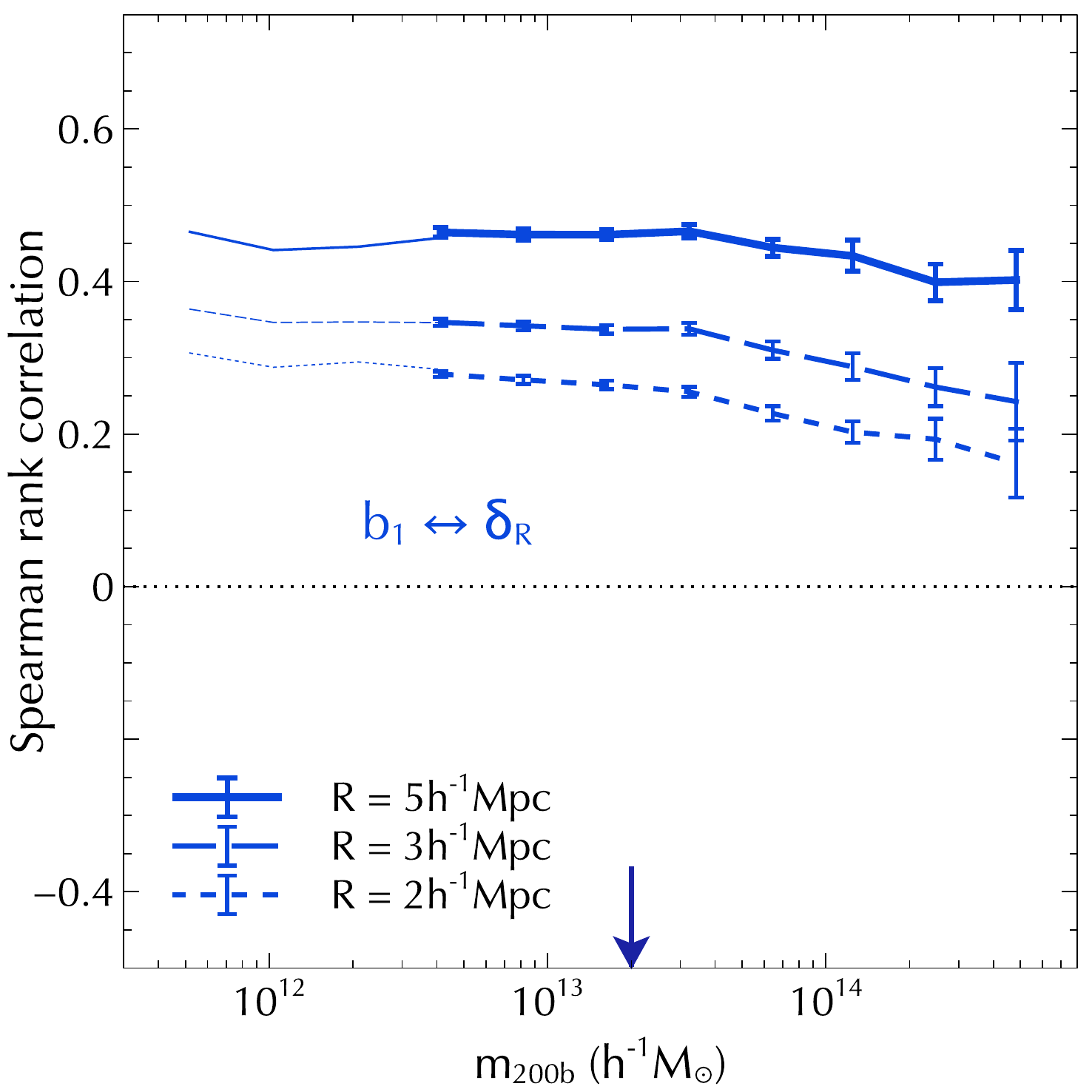}
\caption{\emph{(Left panel:)} Halo-by-halo bias against halo mass for individual haloes in one realisation of our simulations randomly downsampled to 20,000 objects, with the points coloured according to $1+\delta_{5\Mpch}$, i.e., the halo-centric overdensity Gaussian-smoothed on scale $5\Mpch$. The symbols joined by solid lines show the median bias for four bins of $1+\delta_{5\Mpch}$ as indicated, as a function of mass. We averaged the median measurements over \NSIM\ realisations of the default box and the error bars show the standard error over these realisations. We see that the median bias at fixed overdensity depends only weakly on halo mass. \emph{(Right panel:)} Spearman rank correlation coefficient between $b_1$ and $\delta_R$ for $R=2,3,5\Mpch$ as indicated (thick lines), averaged over \NSIM\ realisations of the default box, with the error bars showing the standard error on the mean. The thin lines extending to low masses show the results of the single high resolution box; the corresponding trends are simple extrapolations of those in the default box, showing that the results are numerically converged. The arrow in each panel marks the characteristic mass $m_\ast$ obtained from the peak of the halo mass distribution function.}
\label{fig:b1Vsmass-delta}
\end{figure*}

\section{Large scale halo bias and small scale environment}
\label{sec:bias-env}
\noindent
Much of the analysis below will involve studying the connection between halo clustering at large scales and the tidal environment of the haloes at small or intermediate scales. Our main goal in this section is to identify a convenient variable that quantifies the tidal anisotropy of the immediate halo environment, whose correlation with the \emph{large scale} environment can then be a probe of assembly bias. As a warm-up, let us first explore the relation between the halo-by-halo bias variable $b_1$ defined above and the simplest variable that characterises halo environment, namely, the dark matter density contrast $\delta_R$ smoothed on some fixed large scale $R$.

\subsection{Correlation between $b_1$ and $\delta_R$}
\label{subsec:b1<->deltaR}
\noindent
In Figure~\ref{fig:hist-delta5Mpch}, we show the distribution of $1+\delta_R$ centered on haloes and smoothed with a Gaussian window of radius $R=5\Mpch$. This was done by first computing $\delta$ on a $N_{\rm g}=512^3$ grid using CIC interpolation and smoothing in Fourier space (i.e., multiplying $\delta(\mathbf{k})$ with $\e{-k^2R^2/2}$), and then transforming back to real space and interpolating the smoothed field to the locations of the haloes to get $\delta_R$. We used all haloes in a realisation that passed the cuts discussed in section~\ref{sec:numerics} and averaged over \NSIM\ realisations of the default box. 
We have split the distribution in Figure~\ref{fig:hist-delta5Mpch} as arising from four categories -- nodes, filaments, voids and sheets -- determined by the number of positive eigenvalues of the tidal tensor $T_{ij}$ \citep{hahn+07} as described in Appendix~\ref{app:tidal}.

\begin{figure*}
\centering
\includegraphics[width=0.8\textwidth]{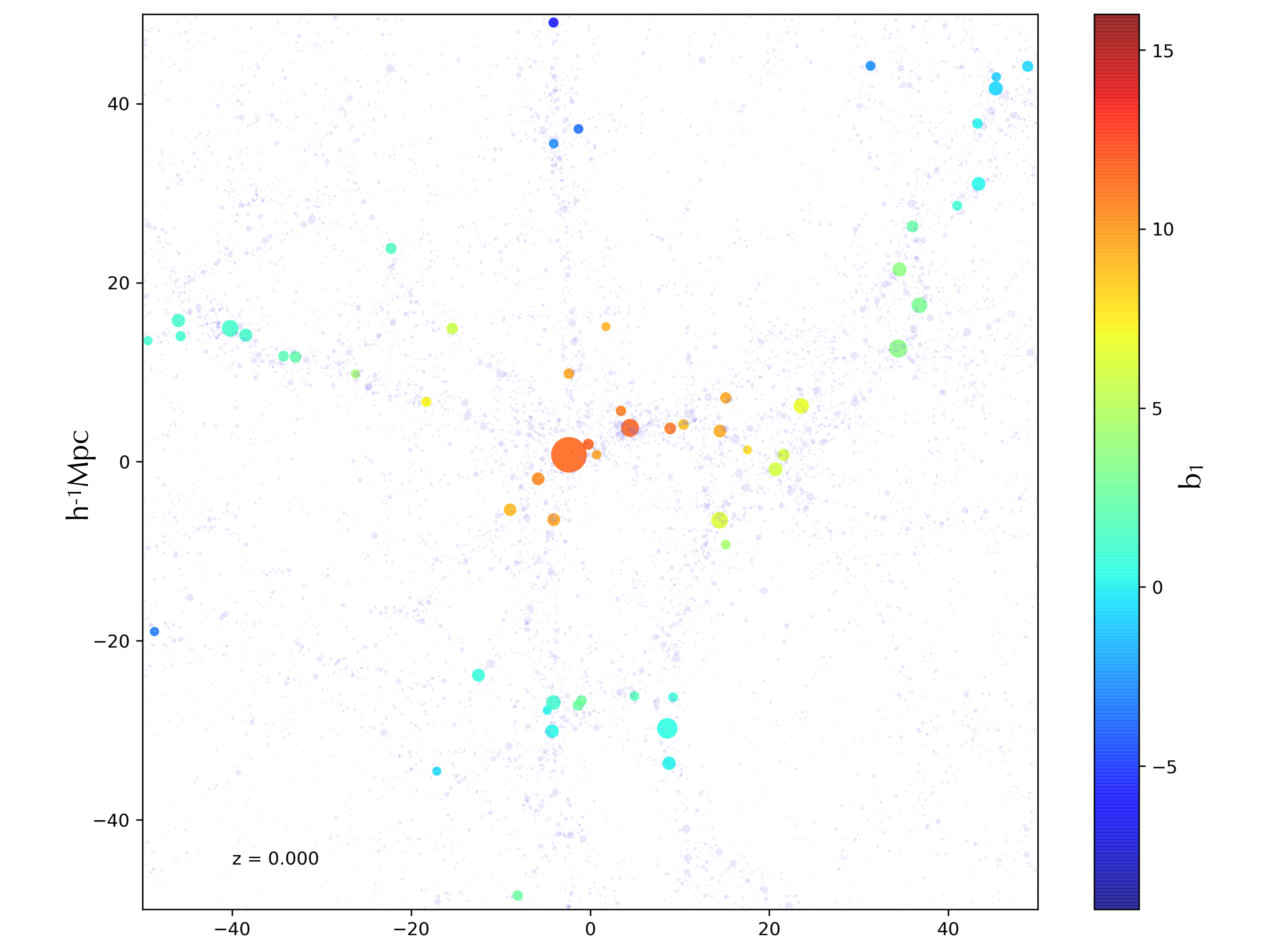}\\
\includegraphics[width=0.8\textwidth]{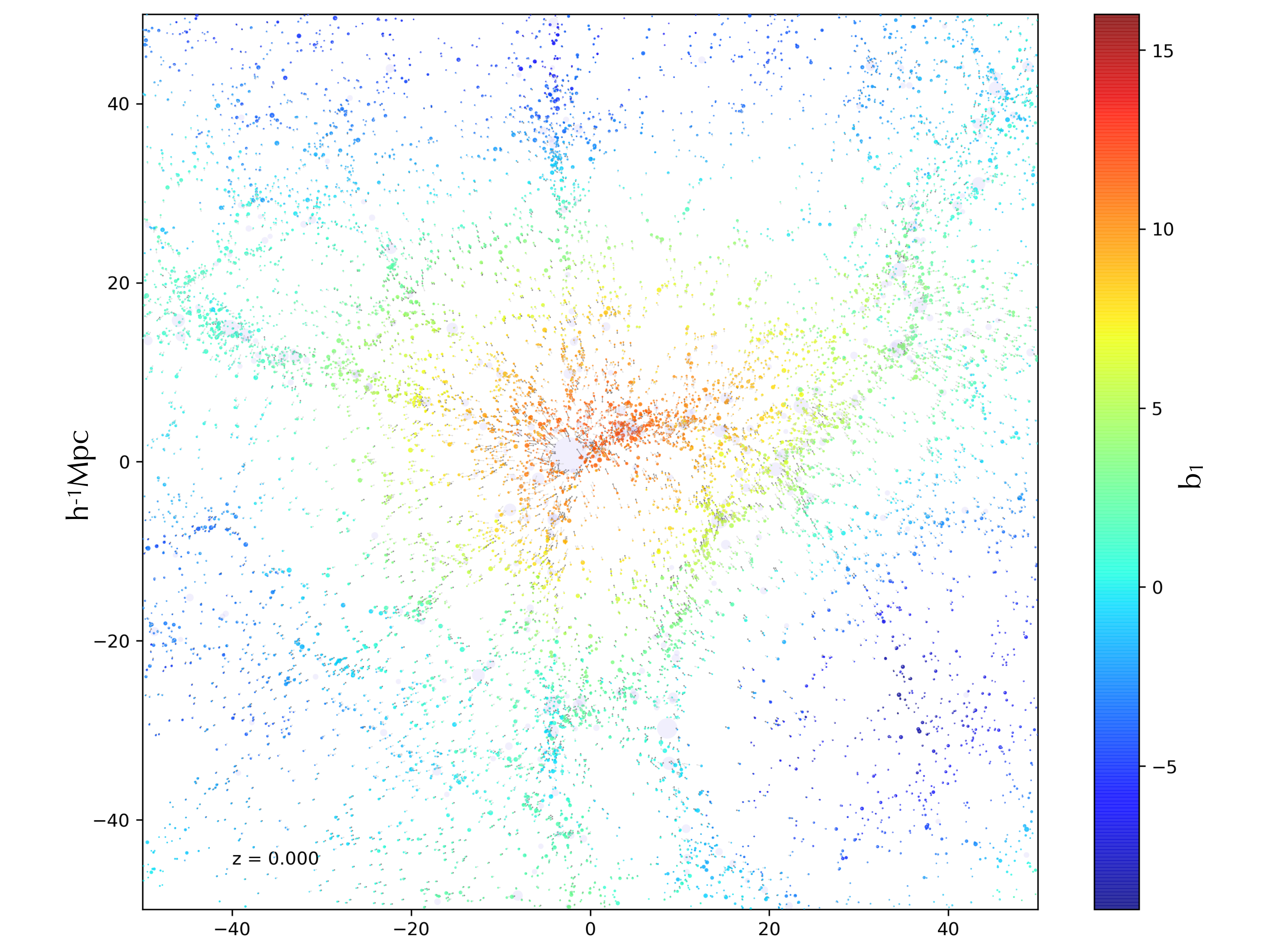}
\caption{Visualisation of haloes in a $100\Mpch\times100\Mpch\times30\Mpch$ volume in the high resolution box, centered on the halo with the largest value of $\alpha_R$ with $R=R_{\rm G,eff}^{(4R_{\rm 200b})}$ (equation~\ref{eq:alpha-def}) and projected along the $30\Mpch$ direction. Circles indicate halo positions, with radius $1.25R_{\rm 200b}$ each (to scale). Opaque coloured circles in the \emph{top panel} correspond to massive haloes with $m>1.5m_\ast$, with the colour indicating the value of $b_1$ for each halo as per the colour bar. Similarly, the \emph{bottom panel} focuses on low mass haloes with $m_{\rm min}<m<m_\ast/4$. 
Arrows on the opaque circles (clearer in the bottom panel) indicate the halo bulk velocity, scaled up to the straight-line distance the halo would travel in $500h^{-1}{\rm Myr}$. 
Transparent blue circles in each panel indicate all haloes with $m>m_{\rm min}$ that are not in the respective bin. 
For this plot we use $m_{\rm min}=9.63\times10^{10}\Mh$ (which is one half of the value used for this box in the main analysis; see section~\ref{sec:numerics}). See main text for a discussion.}
\label{fig:halovisual-b1}
\end{figure*}

The \emph{left panel} of Figure~\ref{fig:b1Vsmass-delta} shows the scatter plot of $b_1$ and mass, coloured by $1+\delta_{5\Mpch}$. There is an obvious correlation visible, with a largely vertical trend in which $b_1$ increases monotonically with $\delta_{5\Mpch}$. 
The symbols with errors show the median bias as a function of mass, in four bins of $\delta_{5\Mpch}$ and averaged over \NSIM\ realisations of the default box. It is clear that, at fixed $\delta_{5\Mpch}$, the trend of bias with halo mass is weak.
This trend is consistent with previous results in the literature, which have shown that large scale bias is more strongly correlated with halo-centric overdensity than it is with halo mass \citep[see, e.g.,][]{as07,ss18}. The \emph{right panel} of the Figure explores this further, showing the Spearman rank correlation coefficient between $b_1$ and $\delta_R$ as a function of halo mass, for $R=2,3,5\Mpch$. We see that the strength of the correlation is only a weak function of mass for each smoothing scale, but monotically increases with $R$. This increase with $R$ is not surprising, since our estimator for $b_1$ itself is ultimately measuring a large scale halo-centric overdensity, so that $b_1$ and $\delta_R$ are measuring essentially the same quantity for large $R$. 
To appreciate this point better, Figure~\ref{fig:halovisual-b1} shows a visualisation of the haloes in a subvolume of our high resolution box, with haloes shown as circles whose radii scale with $R_{\rm 200b}$ and whose colour scales with halo bias $b_1$ as indicated by the colour bar. The panels focus on massive \emph{(top)} and low mass haloes \emph{(bottom)}. 
We discuss some connections between halo-by-halo bias and gravitational redshift measurements \citep[see, e.g.][]{whh11,croft13,alam+17} in Appendix~\ref{app:gravredshift}.
In Appendix~\ref{subapp:b1-delta}, we present analytical arguments that explain the size of the scatter in $b_1$ at fixed mass and also qualitatively reproduce the trends seen in Figure~\ref{fig:b1Vsmass-delta}.

The variable $\delta_R$ at fixed smoothing scale $R$ might seem to be a natural choice for determining environmental properties. We see in Figure~\ref{fig:hist-delta5Mpch}, however, that $\delta_{5\Mpch}$ is not a particularly strong discriminator of environment into nodes, filaments, etc., and we have checked that similar results are true at smaller smoothing scales as well. There is also not much to be gained by smoothing at fixed larger scales, either, since we are ultimately interested in the tidal environment on scales close to the halo size, which would physically correspond to the tidal forces being experienced by individual haloes. We therefore conclude that we should look for variables defined close to the halo size that discriminate between different environments better than $\delta_R$, and also correlate more strongly with $b_1$ than does $\delta_R$.

\subsection{Tidal anisotropy $\alpha_R$}
\label{subsec:alphaR}
\noindent
The rotational invariants of the tidal tensor $T_{ij}$ beyond its trace $\delta_R$ are a natural starting point in looking for discriminatory variables. One such variable $q_R^2$, sometimes referred to in the literature as tidal shear, is particularly promising. This is defined as \citep{hp88,ct96a}
\begin{align}
q_R^2 &\equiv I_1^2 - 3I_2 \notag\\
&= \frac12\left[\left(\lambda_3-\lambda_1\right)^2 + \left(\lambda_3-\lambda_2\right)^2 + \left(\lambda_2-\lambda_1\right)^2\right]\,,
\label{eq:q2-def}
\end{align}
where $\lambda_1\leq\lambda_2\leq\lambda_3$ are the eigenvalues of $T_{ij}$ and $I_1=\lambda_1+\lambda_2+\lambda_3$ and $I_2=\lambda_1\lambda_2+\lambda_2\lambda_3+\lambda_3\lambda_1$ are its first two rotational invariants. A closely related variable $s^2=2q^2/3$ has been used in the recent literature in the context of measuring `non-local' bias \citep{chan/scoccimarro/sheth:2012,baldauf/seljak/etal:2012,Saito+14}. 

For a \emph{Gaussian} random field, the shear $q_R^2$ has the remarkable property that its distribution is \emph{independent} of the trace $\delta_R$ \citep[and can be shown to be Chi-squared with $5$ degrees of freedom, see][]{st02}. In general, $q_R^2$ reflects the anisotropy of the tidal environment at any scale $R$, vanishing for a perfectly isotropic environment. In terms of the more commonly used anisotropy measures  `ellipticity' $e_R\equiv(\lambda_3-\lambda_1)/2\delta_R$ and `prolateness' $p_R\equiv(\lambda_3-2\lambda_2+\lambda_1)/2\delta_R$ \citep[e.g.,][]{bbks86,bm96}, we have\footnote{Note that \citet{bbks86} used the ellipticity and prolateness defined with the Hessian of the overdensity field, $\p_i\p_j\delta_R$, while \citet{bm96} distinguished between these and analogous quantities defined using the tidal tensor $T_{ij}$. The variables we refer to above correspond to the latter; $e_v$ and $p_v$ in the notation of \citet{bm96}.} $q_R^2=\delta_R^2\left(3e_R^2+p_R^2\right)$. So we expect that $q_R^2$ defined close to the halo scale should retain substantial information regarding the tidal anisotropy of the halo environment. 

For the \emph{non-linear} dark matter field, unfortunately, $q_R^2$ is quite strongly correlated with $\delta_R$. To see why this is to be expected, consider that the density contrast in 2LPT can be written in terms of the Gaussian-field $\delta$ and $q^2$ as $\delta_{\rm 2LPT}=\delta + (17/21)\delta^2+(4/21)q^2$. Approximating the nonlinear shear by its value for the Gaussian field then already shows that one might expect the correlation coefficient between $q_R^2$ and $\delta_R$ to be $\simeq0.12\,\sigma\times\left(1+\Cal{O}(\sigma^2)\right)$ at scales where 2LPT is valid, where $\sigma^2=\avg{\delta^2}=\avg{q^2}$, with a stronger correlation at smaller scales. This means that any correlation $q_R$ might have with $b_1$ could easily be contaminated by the correlation between $b_1$ and $\delta_R$, and not necessarily be a measure of anisotropy alone.

After some experimentation, we have found that the following variable has the properties we require for quantifying tidal anisotropy at the halo scale, beyond what is measured by $\delta_R$:
\be
\alpha_R \equiv \left(1+\delta_R\right)^{-1}\sqrt{q_R^2}\,.
\label{eq:alpha-def}
\ee
We demonstrate this next with a series of measurements.
Before we do so, however, it is worth mentioning that we have also explored analogous variables constructed using the Hessian of the density $\p_i\p_j\delta_R$. Indeed, several studies in the past and more recently have attempted to define the large scale environment through the density and its derivatives \citep[see, e.g.][]{aragoncalvo07,sousbie11,yang+17}. We find, however, that these tend to be poorer discriminators of the local web environment than the variables based on the tidal tensor \citep[in agreement with][]{wmjyw11,swm15}. 
We have not explored variables based on the velocity shear $(\p_iv_j+\p_jv_i)/2$ \citep{hahn+09,hoffman+12} which would differ from the tidal tensor due to nonlinear evolution. 
In principle, one might make more objective statements by comparing the utility of variables defined using the tidal tensor, density Hessian or velocity shear using information theoretic criteria such as those proposed by \citet{lljw16}; however, this is beyond the scope of the present work.

\begin{figure}
\centering
\includegraphics[width=0.45\textwidth]{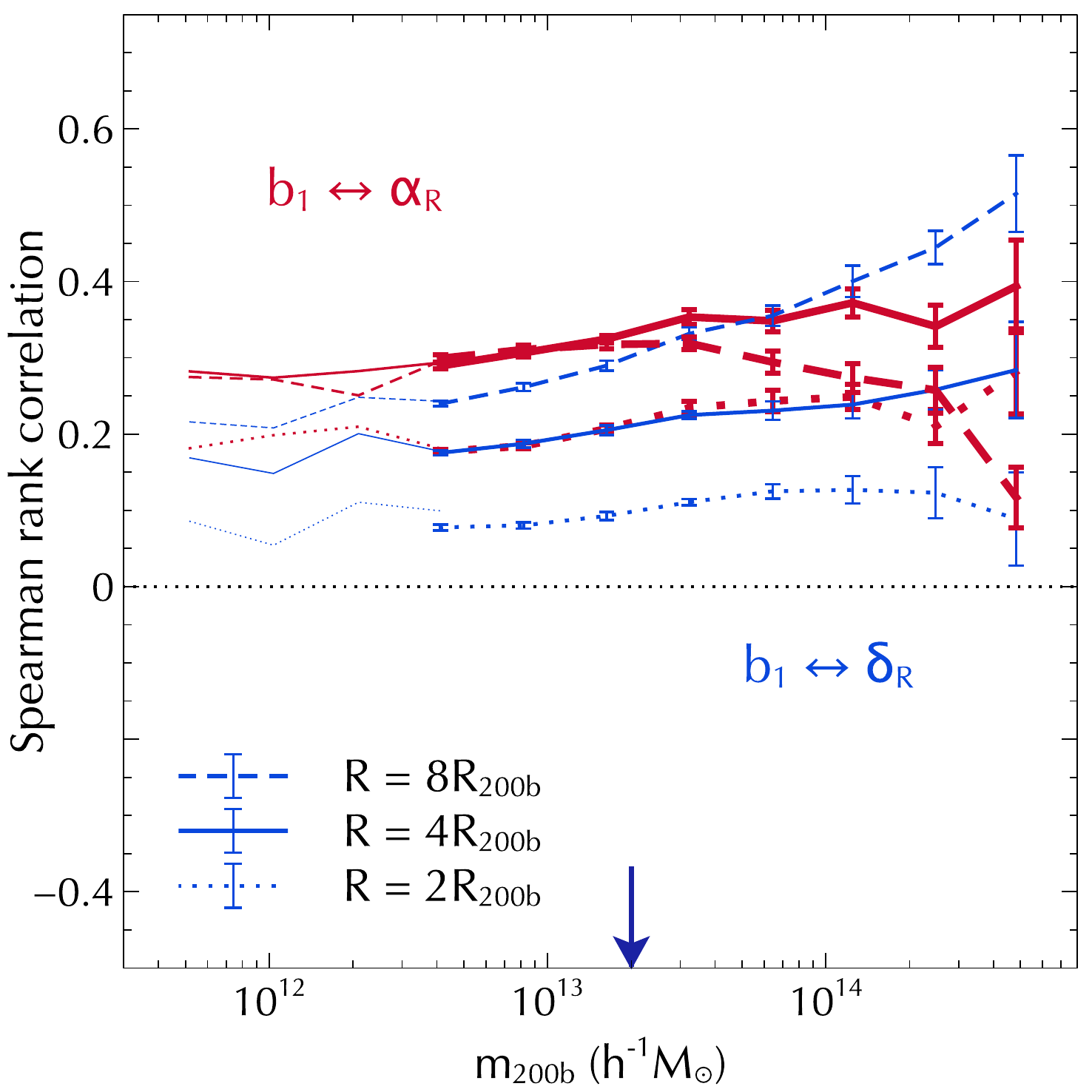}
\caption{Spearman rank correlation coefficient between $b_1$ and $\alpha_R$ (red, thick) for $R=R_{\rm G,eff}^{(2R_{\rm 200b})}$ (dotted), $R_{\rm G,eff}^{(4R_{\rm 200b})}$ (solid) and $R_{\rm G,eff}^{(8R_{\rm 200b})}$ (dashed) averaged over \NSIM\ realisations of the default box, with error bars indicating the error on the mean of the \NSIM\ realisations. For comparison, the correlation between $b_1$ and $\delta_R$ for the same smoothing scales is displayed as the thinner blue curves. The thinnest lines extending to low masses show the corresponding results of the single high resolution box; these are consistent with simple extrapolations of those in the default box. The arrow marks the characteristic mass $m_\ast$ obtained from the peak of the halo mass distribution function.}
\label{fig:rankcorr-248R200b}
\end{figure}

\begin{figure}
\centering
\includegraphics[width=0.45\textwidth]{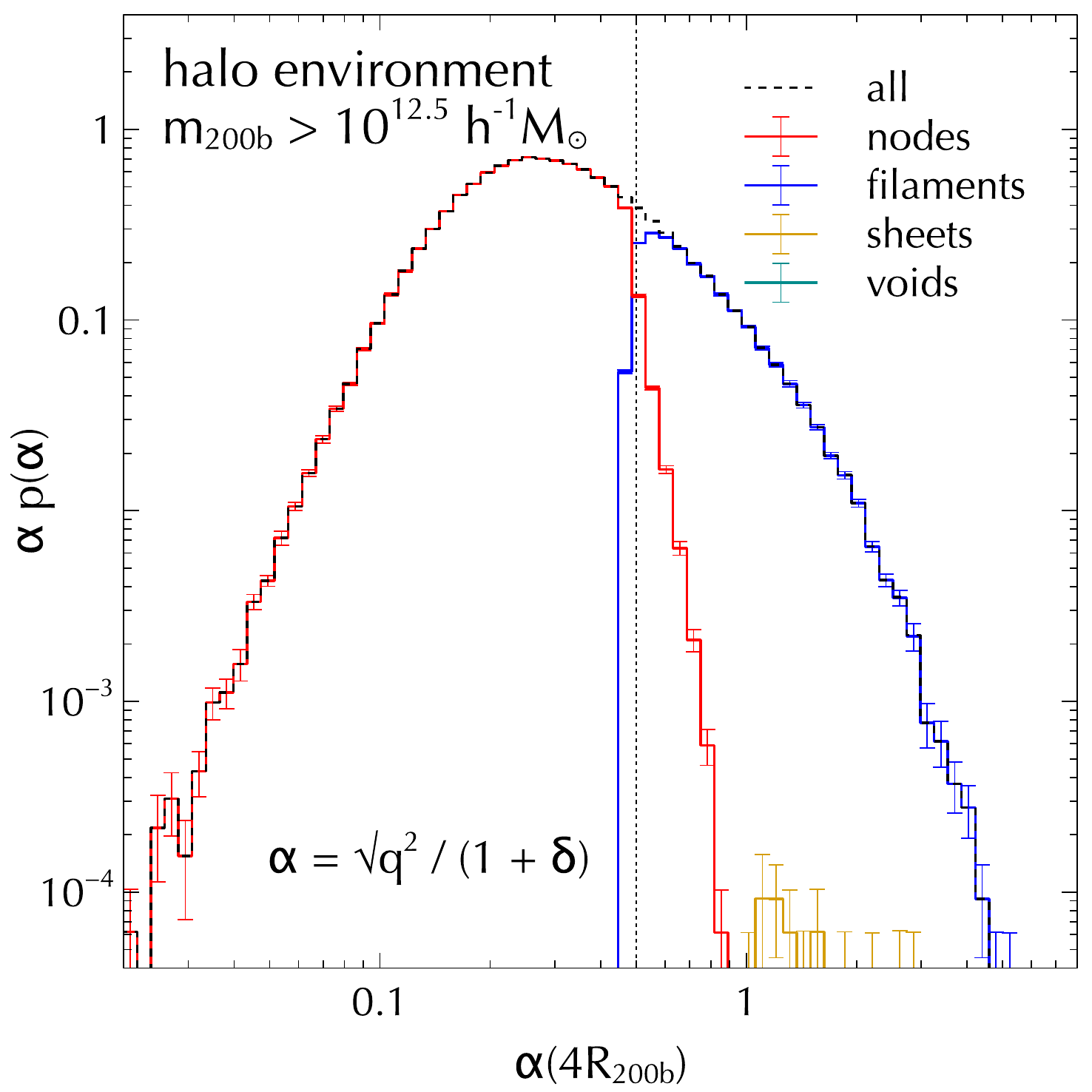}
\caption{Tidal anisotropy in halo environments. Histograms show the distribution of $\alpha_R$ (equation~\ref{eq:alpha-def}) centered on haloes selected as described in the text and smoothed with a Gaussian window of radius $R=R_{\rm G,eff}^{(4R_{\rm 200b})}$ (equations~\ref{eq:RGeff-Mhalo} and~\ref{eq:RGeff-K}), averaged over \NSIM\ realisations of the default box. The error bars indicate the standard error on the mean over the \NSIM\ realisations. Dashed histogram shows the total distribution, while the various colours indicate different halo environments: essentially no sheet or void environments are measured, and there is a sharp distinction between filaments ($\alpha_R\gtrsim0.5$) and nodes ($\alpha_R\lesssim0.5$, compare Figure~\ref{fig:hist-delta5Mpch}). The vertical dotted line indicates $\alpha_R=0.5$.}
\label{fig:hist-alpha4R200b}
\end{figure}

\subsection{Correlation between $b_1$ and tidal anisotropy}
\label{subsec:b1<->alphaR}
\noindent
Figure~\ref{fig:rankcorr-248R200b} shows the Spearman rank correlation between $b_1$ and $\alpha_R$ (red, thick) as a function of halo mass, smoothed at the Gaussian equivalent of $2R_{\rm 200b}$ (dotted), $4R_{\rm 200b}$ (solid) and $8R_{\rm 200b}$ (dashed) (see equations~\ref{eq:RGeff-Mhalo} and~\ref{eq:RGeff-K})\footnote{For reference, for $m_{\rm 200b}=\{10^{11.6},10^{12.5},10^{13.3}\}\Mh$ (corresponding to the minimum mass thresholds for our two boxes and the characteristic mass for our chosen cosmology), we have $R_{\rm G,eff}^{(4R_{\rm 200b})}=\{0.328,0.655,1.21\}\Mpch$, respectively.}. For comparison, we also display the corresponding correlation between $b_1$ and $\delta_R$ (blue, thin) at each smoothing scale (c.f. Figure~\ref{fig:b1Vsmass-delta}). For $2R_{\rm 200b}$ and $4R_{\rm 200b}$, we see that there is a statistically significant \emph{positive} correlation between $b_1$ and $\alpha_R$, which is \emph{stronger} than the corresponding correlation between $b_1$ and $\delta_R$.\footnote{Notice that, had we set $R$ to be the equivalent of $R_{\rm 200b}$, we would expect essentially \emph{no} correlation between $\delta_R$ and $b_1$, since the former would be simply $\simeq199$ for \emph{every} halo.} Also, the $b_1\leftrightarrow\alpha_R$ correlation is stronger at $4R_{\rm 200b}$ than at $2R_{\rm 200b}$. At $8R_{\rm 200b}$, on the other hand, we see that (a) the $b_1\leftrightarrow\alpha_R$ correlation is generally \emph{weaker} than at $4R_{\rm 200b}$ and (b) the correlation between $\delta_R$ and $b_1$ is generally stronger than that between $\alpha_R$ and $b_1$. In Appendix~\ref{subapp:4R200b}, we argue that the size of the sphere around the halo that is currently decoupling from the Hubble flow and turning around is likely to be close to $4$-$6R_{\rm 200b}$, which might plausibly explain why the tidal anisotropy on this scale shows the strongest correlation with large scale environment.

\begin{figure}
\centering
\includegraphics[width=0.45\textwidth]{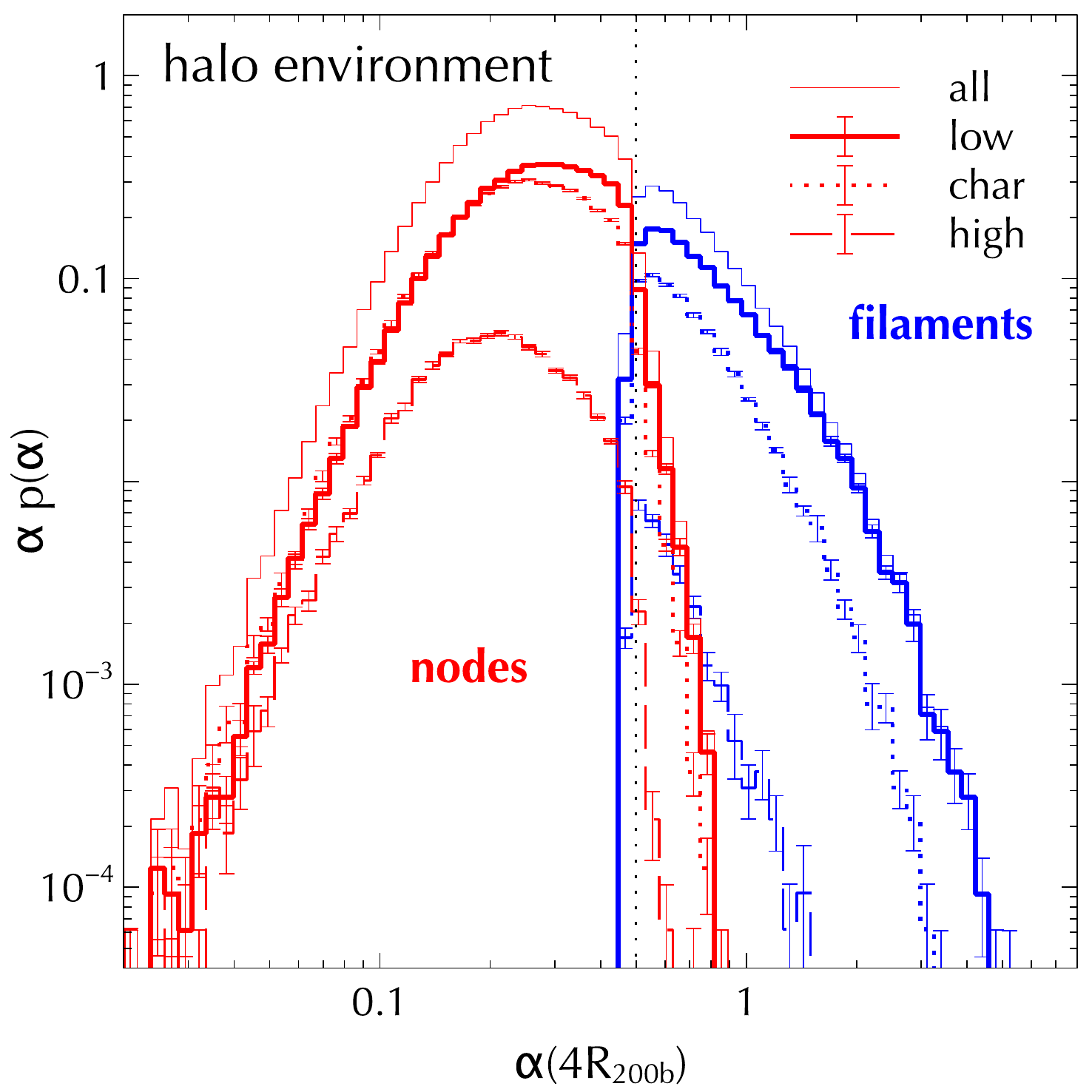}
\caption{Similar to Figure~\ref{fig:hist-alpha4R200b}, showing histograms only for nodes and filaments averaged over \NSIM\ realisations of the default box, with the histograms now split into three bins of halo mass -- \textbf{low}: $m_{\rm min}\leq m < m_\ast/3$ (thick solid); \textbf{char}: $m_\ast/3\leq m<2m_\ast$ (dotted); \textbf{high}: $2m_\ast\leq m<m_{\rm max}$ (long dashed), where $m_{\rm min}$ and $m_{\rm max}$ were defined in section~\ref{sec:numerics}. Thin solid histograms show the respective sums over the three bins for nodes and filaments. The vertical dotted line indicates $\alpha_R=0.5$.}
\label{fig:alpha4R200b-mass}
\end{figure}

These results support our claim that this variable, when defined close to the halo scale (we will use the Gaussian equivalent of $4R_{\rm 200b}$ hereon), is a better indicator than $\delta_R$ of the relation between $b_1$ and the degree of tidal anisotropy around haloes.
To further establish the usefulness of $\alpha_R$ at $R=R_{\rm G,eff}^{(4R_{\rm 200b})}$, we explore the behaviour of histograms of $\alpha_R$ in different web environments in Figure~\ref{fig:hist-alpha4R200b}, which is formatted similarly to Figure~\ref{fig:hist-delta5Mpch} and shows the distribution of $\alpha_R$ for haloes with $N_{\rm p}^{\rm (halo)}\geq1600$ averaged over \NSIM\ realisations of the default box. At these scales, essentially \emph{no} halo is classified as being in a sheet or void, which is easy to understand if we consider that, as $R\to R_{\rm 200b}$, the immediate environment of a halo must be dominated by infall of matter onto the halo. We clearly see that $\alpha_R$ distinguishes quite sharply between traditionally defined filament and node environments, with $\alpha_R \gtrsim 0.5$ ($\alpha_R\lesssim 0.5$) corresponding to filaments (nodes). 
We emphasize, however, that the continuous variable $\alpha_R$ gives us more flexibility in exploring tidal anisotropy than does the traditional filament/node split. Although we will loosely refer to values of $\alpha_R$ above and below $0.5$ as filament-like and node-like, respectively, our $\alpha_R$-based analysis below does not treat $\alpha_R=0.5$ as special in any way. In fact, we will see later that a more useful notion of transition between anisotropic and isotropic environments occurs around $\alpha_R\simeq0.2$, something that would be missed by the traditional node/filament definition.

\begin{figure}
\centering
\includegraphics[width=0.45\textwidth]{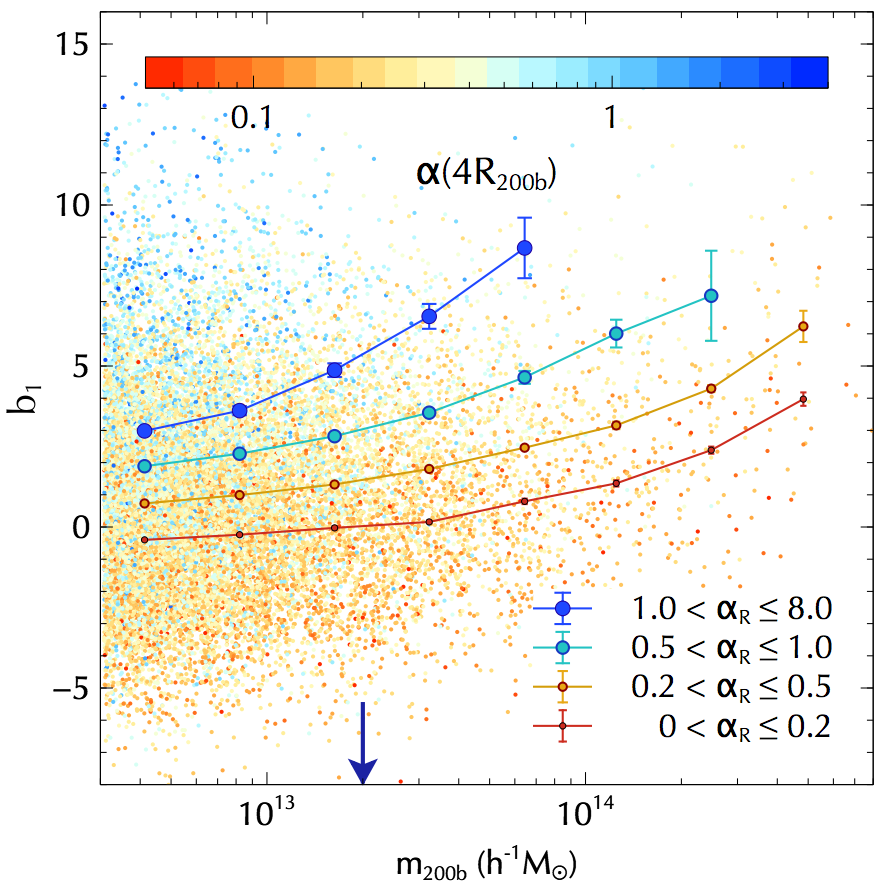}
\caption{Halo-by-halo bias against halo mass, with the points showing measurements of individual haloes in one realisation of our default box, randomly downsampled to 20,000 objects. The points are coloured according to $\alpha_R$ (equation~\ref{eq:alpha-def}) smoothed with a Gaussian window of radius $R=R_{\rm G,eff}^{(4R_{\rm 200b})}$ (equations~\ref{eq:RGeff-Mhalo} and~\ref{eq:RGeff-K}). 
The symbols with errors connected by solid lines show the median bias as a function of mass, for four bins of $\alpha_R$, averaged over \NSIM\ realisations of the default box (the errors show the scatter around the mean). We see strong trends of halo bias with both $\alpha_R$ as well as halo mass (c.f. the left panel of Figure~\ref{fig:b1Vsmass-delta}).
The arrow marks the characteristic mass $m_\ast$ obtained from the peak of the halo mass distribution function.}
\label{fig:b1-mass-alpha4R200b}
\end{figure}

\begin{figure*}
\centering
\includegraphics[width=0.8\textwidth]{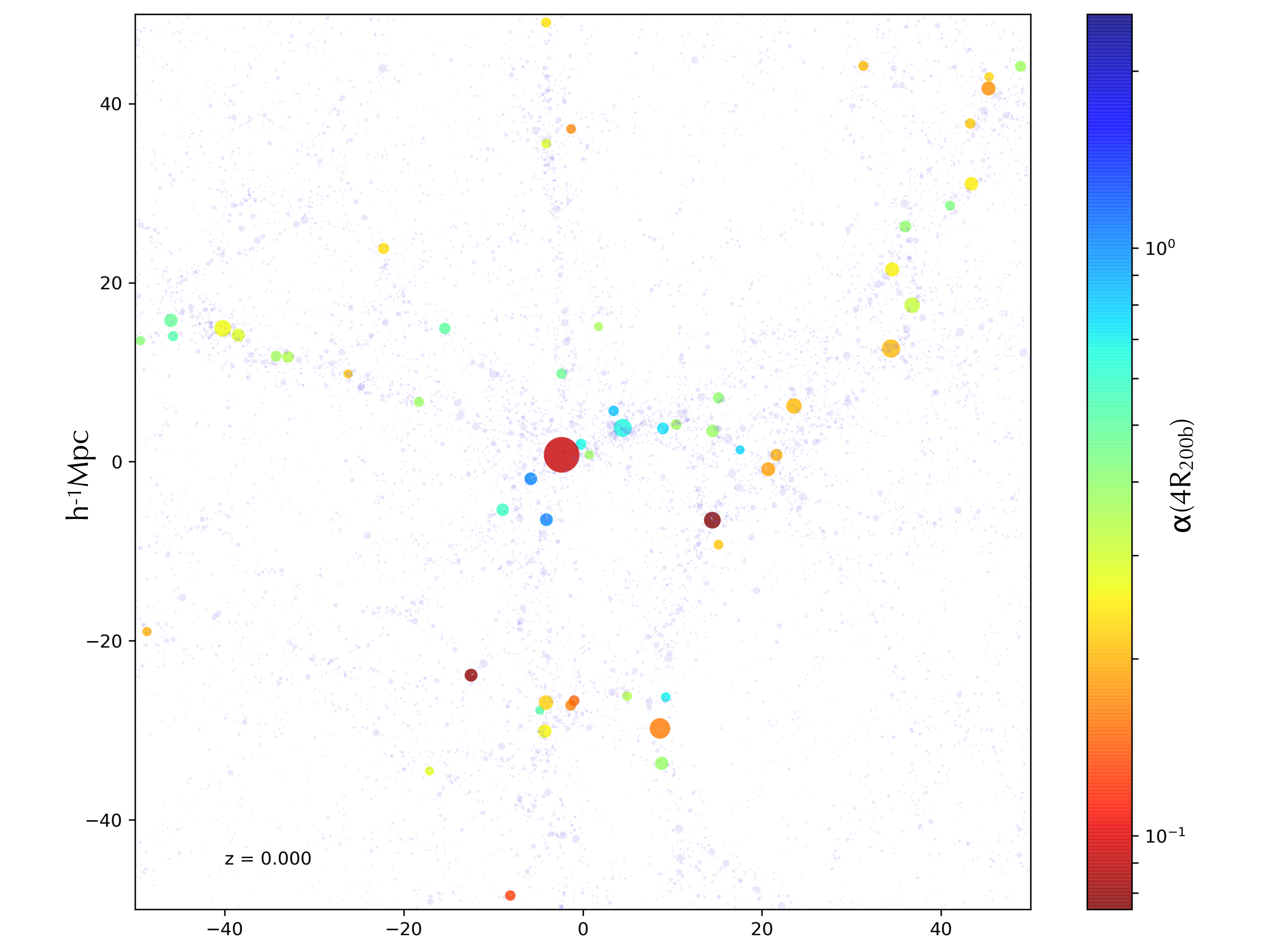}\\
\includegraphics[width=0.8\textwidth]{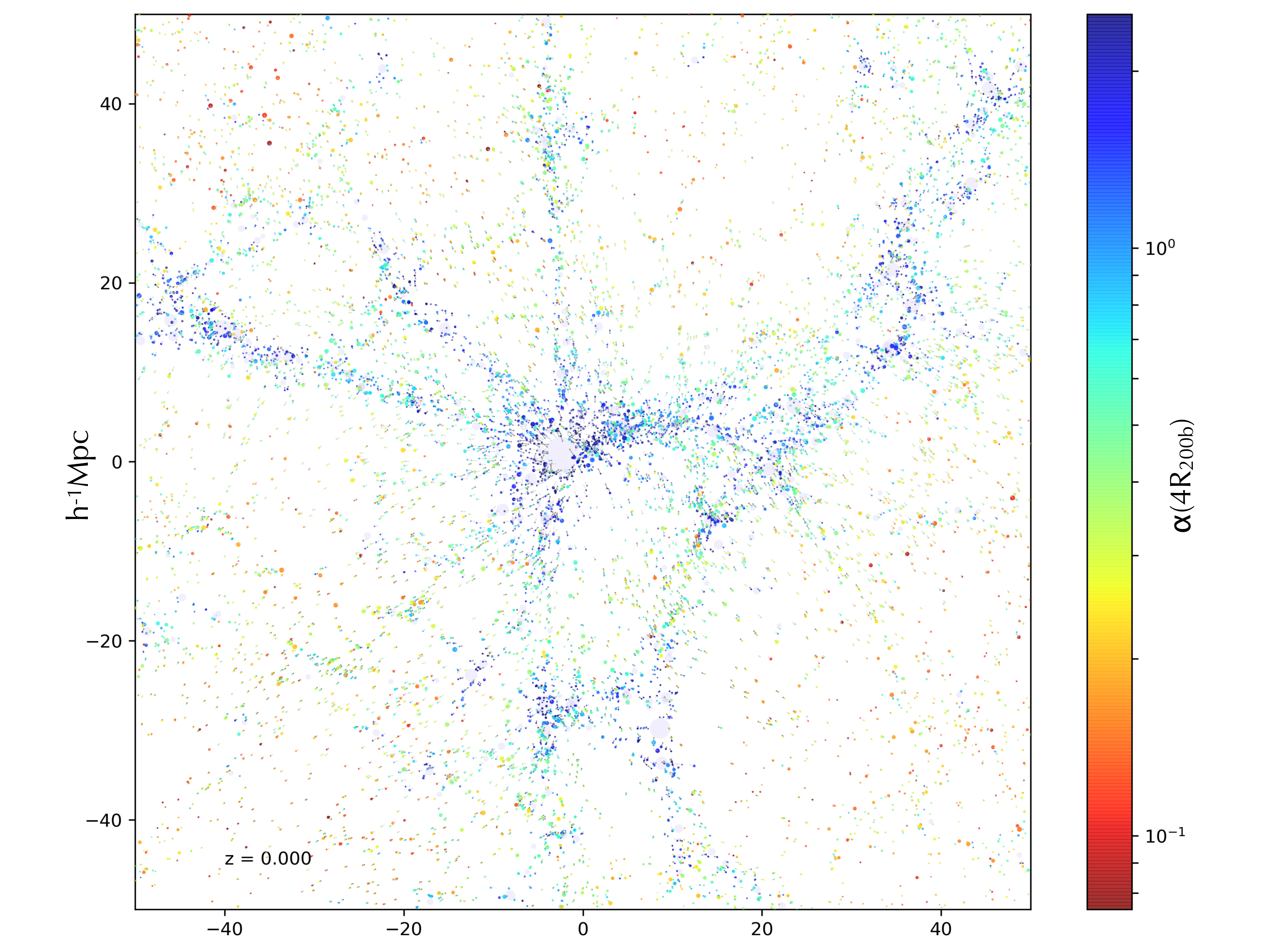}
\caption{Similar to Figure~\ref{fig:halovisual-b1}, but with the colour of the halo markers scaling logarithmically according to the value of $\alpha_R$ with $R=R_{\rm G,eff}^{(4R_{\rm 200b})}$ for each halo as indicated by the colour bar.
Haloes classified as being in anisotropic local environments ($\alpha_R\gtrsim0.5$, blueish colours), particularly the low-mass haloes in the \emph{bottom panel}, clearly trace out large-scale filaments. Haloes with $\alpha_R\lesssim0.2$ (reddish colours), on the other hand, are associated with either dense clusters (massive haloes in the \emph{top panel}) or underdense void-like regions (low-mass haloes in the \emph{bottom panel}). Low-mass haloes in the most anisotropic environments are predominantly associated with nearby massive haloes that generate strong tidal effects in their vicinity.}
\label{fig:halovisual}
\end{figure*}

We dissect the behaviour of $\alpha_R$ with halo mass in Figure~\ref{fig:alpha4R200b-mass}, which is similar to Figure~\ref{fig:hist-alpha4R200b}, with the histograms now split into three bins of halo mass, with masses substantially below, close to and substantially above the characteristic mass $m_\ast$, respectively. We clearly see that filamentary haloes are dominated by the lowest mass bin. To see that this is not simply a consequence of the lowest mass bin being the most populous, note that the combined distribution of node+filament haloes in each mass bin forms an envelope whose median decreases from low to high masses. On the other hand, the transition between traditional node and filament environments remains sharp and fixed at $\alpha_R\simeq0.5$. Together, this makes the \emph{fraction} of haloes in any mass bin that are traditionally classified as being in filaments \emph{decrease} with increasing halo mass. The results of our high resolution box (not shown) are qualitatively consistent with these, with the distribution of $\alpha_R$ at the lowest masses extending to somewhat larger values.

Finally, we explore the correlation between $b_1$ and $\alpha_R$ at different halo masses in Figure~\ref{fig:b1-mass-alpha4R200b}, which is similar to the left panel of Figure~\ref{fig:b1Vsmass-delta}, with the points now coloured by $\alpha_R$ defined at $4R_{\rm 200b}$. There is a clear indication that the highest values of $b_1$ at $m<m_\ast$ in the left panel arise predominantly from haloes in filaments \citep[see also][]{bprg17}. 
This is further emphasized by the symbols with error bars joined by solid lines, which show the median bias as a function of halo mass in four bins of $\alpha_R$. There is a clear strong monotonic trend of halo bias with $\alpha_R$. Unlike Figure~\ref{fig:b1Vsmass-delta}, however, there is now a substantial trend of bias with mass even at fixed $\alpha_R$, which reiterates the point that $\alpha_R$ encodes different information about the large scale environment of haloes than variables such as halo mass or $\delta_{5\Mpch}$.
We will return to the dependence of bias on $\alpha_R$ when discussing assembly bias below. 

Figure~\ref{fig:halovisual} is similar to Figure~\ref{fig:halovisual-b1}, except that the colour of the halo markers scales with the tidal anisotropy $\alpha_R$ defined at $R=R_{\rm G,eff}^{(4R_{\rm 200b})}$.
Keeping in mind the histograms in Figure~\ref{fig:alpha4R200b-mass}, we see that haloes with green to blue colours ($\alpha_R\gtrsim0.5$) are classified as being in filament environments at the halo scale, while yellow to red colours correspond to node environments. It is then clear from the bottom panel that low mass haloes classified as being in filaments do, in fact, visually trace out filamentary structures and also predominantly occur in the vicinity of massive objects (which are themselves classified as being in nodes at their correspondingly larger smoothing scale). And low mass haloes far from any massive haloes are predominantly classified as being in nodes. We will see later that the distinction more relevant for assembly bias in fact occurs at smaller values of tidal anisotropy, $\alpha_R\simeq0.2$.

\begin{figure}
\centering
\includegraphics[width=0.45\textwidth]{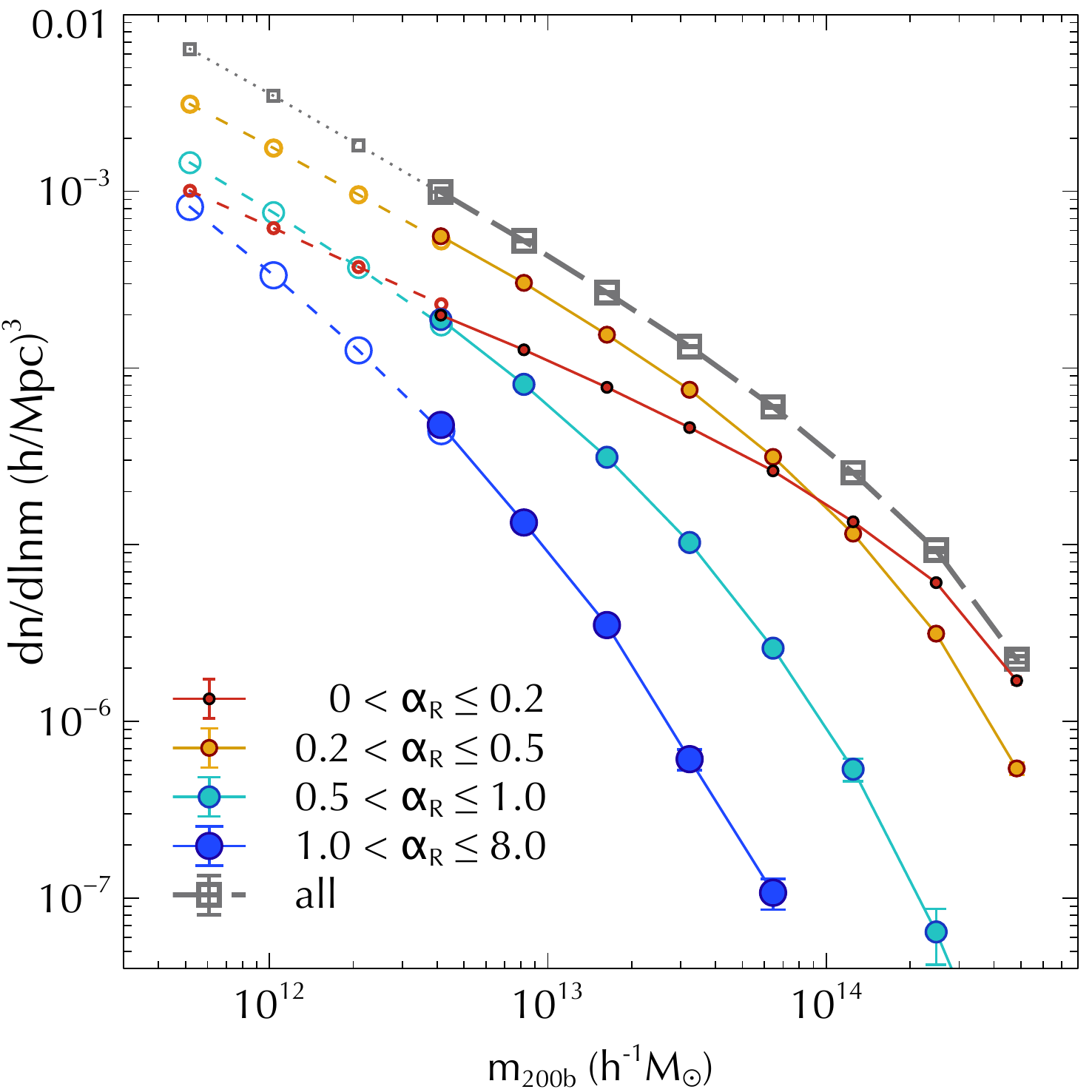}
\caption{Halo mass function in bins of $\alpha_R$ for $R=R_{\rm G,eff}^{(4R_{\rm 200b})}$ averaged over \NSIM\ realisations of our default box (filled circles, with error bars indicating the error on the mean in \NSIM\ realisations). For comparison, the empty gray squares show the mass function of all haloes. The empty symbols joined by thin lines extending to low masses show the results of the single high resolution box; these are consistent with simple extrapolations of the measurements in the default box.}
\label{fig:dndlnm-alpha4R200b}
\end{figure}

\subsection{Halo properties and tidal environment}
\label{subsec:haloproptidal}
\noindent
Before turning to a detailed study of assembly bias in different tidal environments, in this section we briefly discuss the variation of halo abundances and halo concentration with the tidal anisotropy $\alpha_R$.

\begin{figure}
\centering
\includegraphics[width=0.45\textwidth]{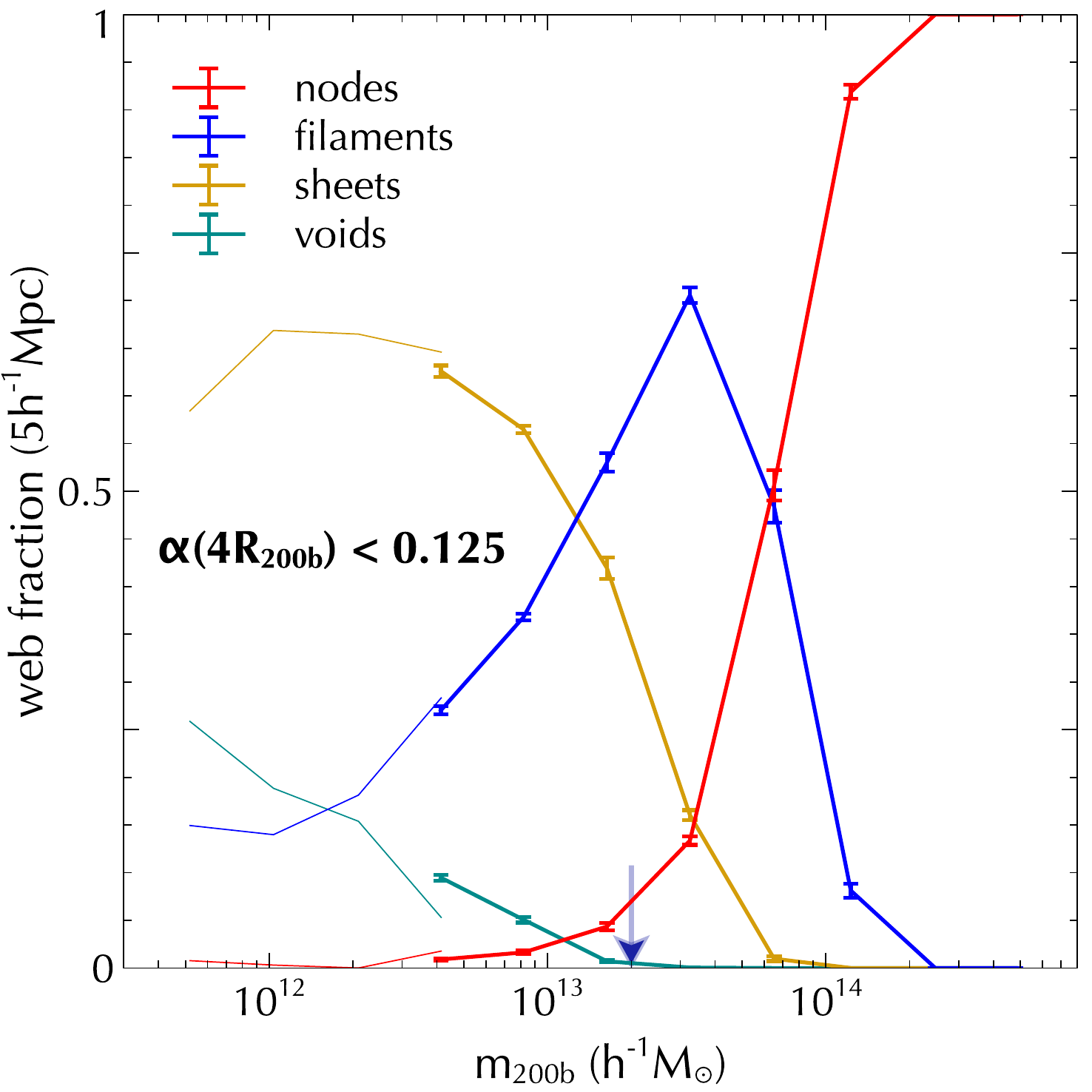}
\caption{Fraction of haloes residing in $5\Mpch$ node, filament, sheet and void environments (distributions favouring higher to lower masses, respectively). Haloes were selected to be those in the most isotropic \emph{local} environments, satisfying $\alpha_R<0.125$ for $R=R_{\rm G,eff}^{(4R_{\rm 200b})}$. Results are averaged over \NSIM\ realisations of our default box, with error bars indicating the error on the mean in \NSIM\ realisations. The thin lines extending to low masses show the results of the single high resolution box; these clearly continue the trends seen in the default box. The arrow marks the characteristic mass $m_\ast$ obtained from the peak of the halo mass distribution function.}
\label{fig:isowebfracn}
\end{figure}

Figure~\ref{fig:dndlnm-alpha4R200b} shows the halo mass function of all haloes (gray squares) and of haloes split into four bins of $\alpha_R$ for $R=R_{\rm G,eff}^{(4R_{\rm 200b})}$ with edges given by $\alpha_R=\{0.0,0.2,0.5,1.0,8.0\}$ (filled circles, size increases with $\alpha_R$). The bins with the two higher $\alpha_R$  values therefore approximately correspond to filamentary environments according to the standard classification, c.f. Figure~\ref{fig:alpha4R200b-mass}). We see that the mass function steadily moves to smaller characteristic masses as $\alpha_R$ increases beyond $\gtrsim0.2$. The mass function in highly isotropic environments with $0<\alpha_R<0.2$, however, dominates only at high masses and falls below that in anisotropic environments at smaller masses. 
This can be understood using Figure~\ref{fig:isowebfracn}, which shows the fraction of haloes residing in $5\Mpch$ node, filament, sheet and void environments, where the haloes were selected to be those in the most isotropic \emph{local} environments, satisfying $\alpha_R<0.125$ for $R=R_{\rm G,eff}^{(4R_{\rm 200b})}$. As expected, at high masses these haloes continue to be classified as being in large scale node environments. At the lowest masses, on the other hand, most of these locally isotropic haloes live in large scale sheet and void environments. The low $\alpha_R$ mass function is therefore a combination of the mass function in dense clusters and sheets/voids. 

\begin{figure}
\centering
\includegraphics[width=0.45\textwidth]{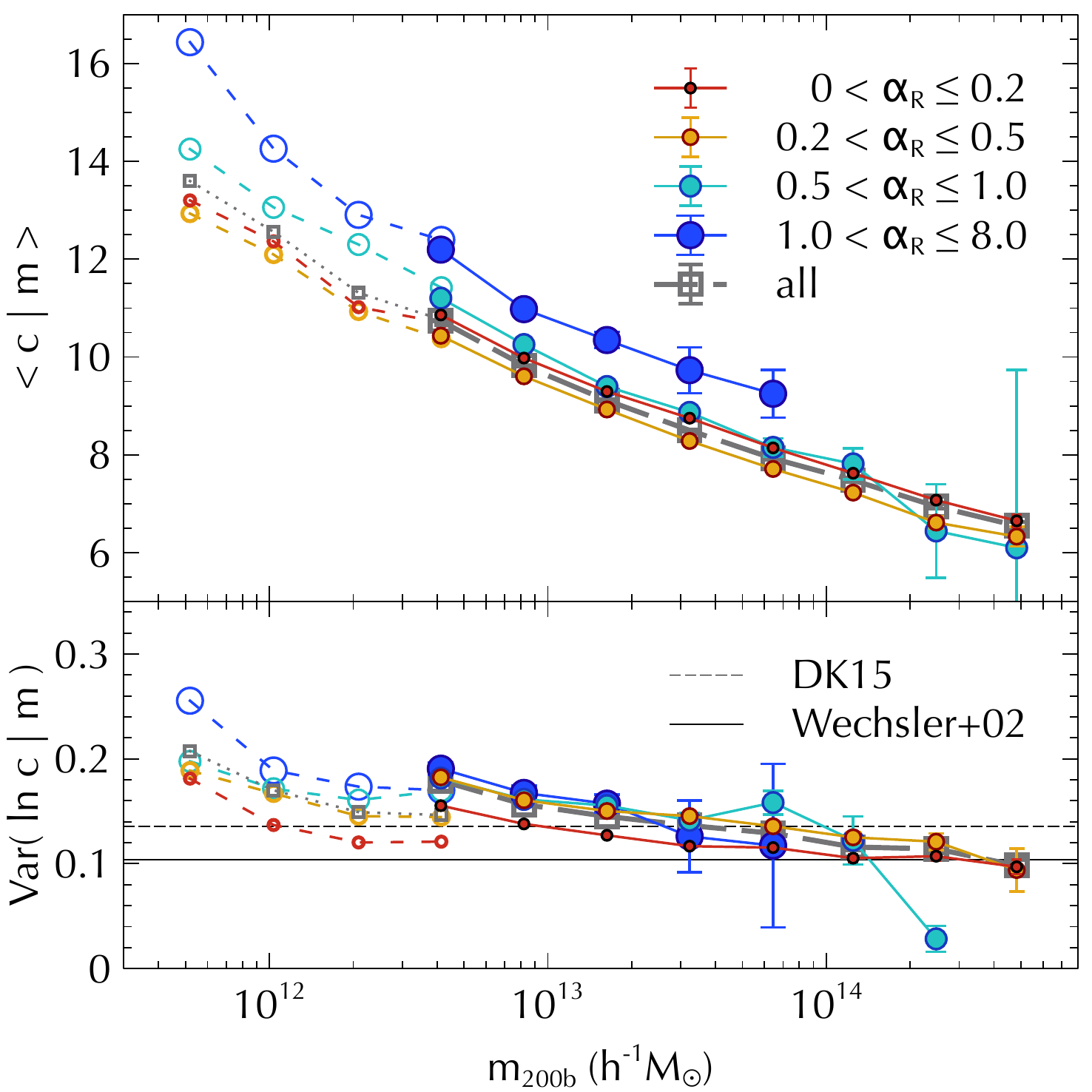}
\caption{Concentration-mass relation for different bins of $\alpha_R$ with $R=R_{\rm G,eff}^{(4R_{\rm 200b})}$. We define halo concentration as $c\equiv c_{\rm 200b}=R_{\rm 200b}/r_{\rm s}$ where $r_{\rm s}$ is the NFW scale radius of the halo. \emph{(Top panel:)} Median concentration as a function of mass $m_{\rm 200b}$, in different tidal environments as defined by four ranges of $\alpha_R$ values as shown, with $R=R_{\rm G,eff}^{(4R_{\rm 200b})}$, formatted identically to Figure~\ref{fig:dndlnm-alpha4R200b}. While the median $c$ in each tidal environment monotonically decreases with mass, there is a non-monotonic trend between $c$ and $\alpha_R$ at fixed mass: $c$ decreases as $\alpha_R$ increases from $0$ to $\sim 0.2$-$0.5$ (node-like environments), and then increases as $\alpha_R$ increases beyond $0.5$ towards more filamentary environments. 
\emph{(Bottom panel:)} Variance of $\ln c$ as function of mass in different tidal environments. This quantity shows weaker trends with mass and tidal anisotropy. For comparison, the solid and dashed horizontal lines show the constant values reported by \citet{wechsler+02} and \citet[][DK15]{dk15}, respectively.
Results in each panel are averaged over \NSIM\ realisations of our default box, with error bars indicating the error on the mean in \NSIM\ realisations. The thin lines extending to low masses show the results of the single high resolution box; these are consistent with extrapolations of the trends seen in the default box, except for some small offsets in the bottom panel.}
\label{fig:concmass-alpha4R200b}
\end{figure}

Defining halo concentration as $c_{\rm 200b}=R_{\rm 200b}/r_{\rm s}$ -- where $r_{\rm s}$ is the scale radius obtained from fitting an NFW profile to the halo mass distribution -- Figure~\ref{fig:concmass-alpha4R200b} shows the median concentration  \emph{(top panel)} and variance of the log-concentration \emph{(bottom panel)} as a function of halo mass in the same four bins of $\alpha_R$ used in Figure~\ref{fig:dndlnm-alpha4R200b}. While the trends in each individual tidal environment are monotonic and qualitatively similar to the result for the full sample, we see a distinct and non-monotonic behaviour of the median concentration as a function of $\alpha_R$, and a weak monotonic dependence of the variance of the log-concentration on $\alpha_R$. The non-monotonicity of the median concentration with $\alpha_R$, which achieves a minimum for $0.2\lesssim\alpha_R\lesssim0.5$, is particularly interesting and would not have been noticed had we used the traditional web-classification for tidal environment (which would club together all haloes with $\alpha_R<0.5$  as being in filamentary environments). The qualitative behaviour of concentration with $\alpha_R$ is also evidently independent of halo mass, which further supports the idea that tidal anisotropy acts as an independent variable determining halo properties.

\section{Assembly bias}
\label{sec:assemblybias}
\noindent
As discussed in the Introduction, it is interesting to theoretically explore the nature of assembly bias and the role played by the tidal environment of haloes in determining the sign and strength of the correlation between internal halo properties and their large scale clustering \citep{hahn+09}. We do this below using our halo-by-halo bias estimator $b_1$.

\subsection{Traditional estimates}
\label{subsec:tradest}
\begin{figure}
\centering
\includegraphics[width=0.45\textwidth]{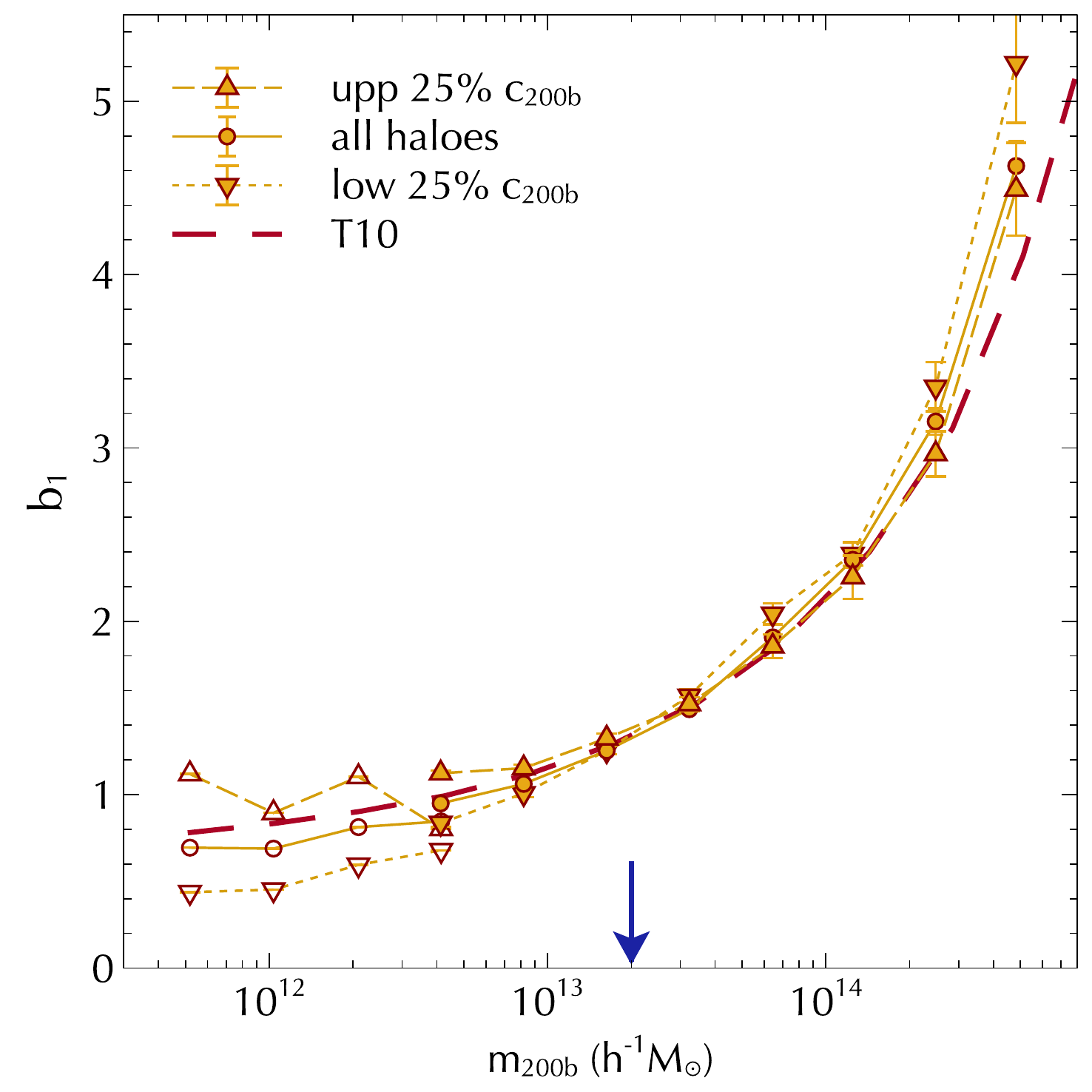}
\caption{Traditional estimate of assembly bias, recovered by binning $b_1$ in mass bins and splitting haloes by concentration quartiles as indicated. Filled symbols show the mean over \NSIM\ realisations of the default box and error bars indicate the standard error on the mean. The empty symbols joined by thin lines extending to low masses show the results of the single high resolution box. The arrow marks the characteristic mass $m_\ast$ obtained from the peak of the halo mass distribution function. The small offset between the results of the default and high resolution boxes is almost certainly a volume effect, since the high resolution box cannot probe the small values of $k$ required for an accurate estimate of $b_1$. We see the well known trend that, at high masses, low concentration haloes are more strongly clustered than high concentration ones, while the inverse is true at low masses.}
\label{fig:tradnalassemblybias}
\end{figure}

\noindent
Our definition of halo-by-halo bias $b_1$ allows us to almost trivially reproduce known results on the large scale clustering of haloes split by any halo property. All that is needed is to calculate the mean value of $b_1$ in appropriately chosen (multi-variate) bins. Focusing for example on halo concentration $c_{\rm 200b}$,  Figure~\ref{fig:tradnalassemblybias} shows the mean bias as a function of halo mass, for all haloes in the mass bin (circles) and for haloes in the upper and lower quartiles of concentration (respectively, upward and downward pointing triangles). We clearly see the well known trend that, at high masses, low concentration haloes are more strongly clustered than high concentration ones, while the trend at low masses is the inverse. The inversion occurs at a mass scale $m_{\rm inv}$ close to the characteristic mass for this cosmology $m_\ast=2\times10^{13}\Mh$ obtained from the peak of the halo mass distribution function and marked by the blue arrow \citep[see][for a discussion of the inversion scale obtained from different techniques]{pp17}. 

\begin{figure*}
\centering
\includegraphics[width=0.85\textwidth]{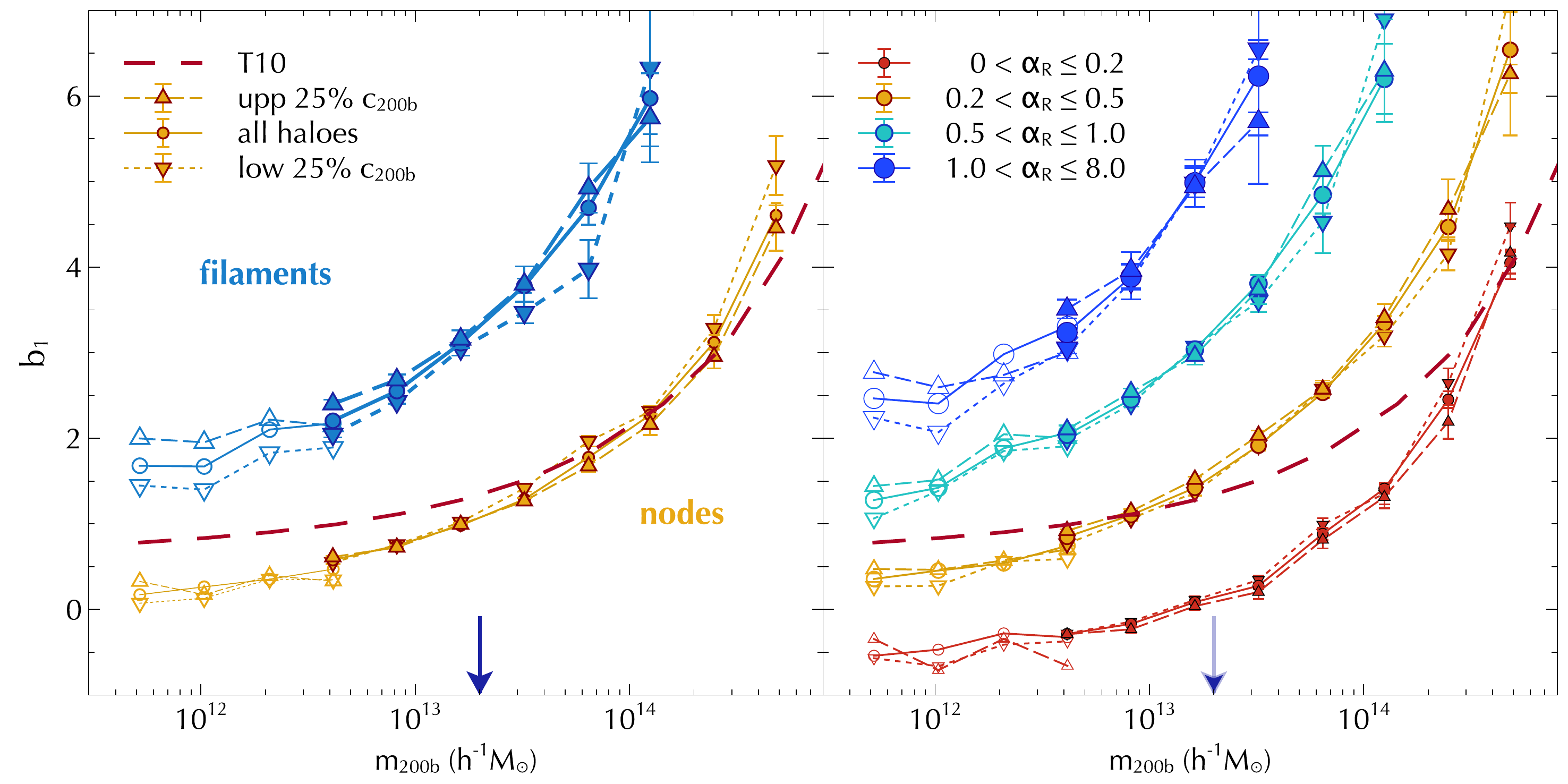}
\caption{Assembly bias in different tidal environments. 
\emph{(Left panel:)} Similar to Figure~\ref{fig:tradnalassemblybias}, but with results shown separately for haloes classified as being in nodes (smaller yellow symbols) and filaments (larger blue symbols) at the Gaussian equivalent of $4R_{\rm 200b}$. 
\emph{(Right panel:)} Concentration-based assembly bias signature dissected as a function of $\alpha_R$ at $R=R_{\rm G,eff}^{(4R_{\rm 200b})}$. Symbols of increasing size (with colours from red to blue) correspond to bins of increasing $\alpha_R$ as indicated. Formatting of point types (circles and triangles) is identical to that in the left panel. 
Filled symbols in each panel show the mean over \NSIM\ realisations of the default box and error bars indicate the standard error on the mean. The empty symbols joined by thin lines extending to low masses show the results of the single high resolution box. The arrow in each panel marks the characteristic mass $m_\ast$ obtained from the peak of the halo mass distribution function, and smooth dashed curve shows the fitting function for linear bias appropriate for $m_{\rm 200b}$-haloes taken from \citet{Tinker10}. The \emph{left panel} shows that the inverted assembly bias trend at low masses in the full sample in Figure~\ref{fig:tradnalassemblybias} arises largely from haloes classified as being in filamentary local environments, as might be expected from the results of \citet{hahn+09} and \citet{bprg17}. The \emph{right panel} shows that this transition happens smoothly as the tidal anisotropy $\alpha_R$ increases from small values (isotropic environments) to large values (anisotropic environments). See the main text and Figure~\ref{fig:rankcorr-b1s-alpha} for further discussion.}
\label{fig:assemblybias-env}
\end{figure*}

We have also checked that we similarly reproduce previous results when haloes at fixed mass are split by their spin or shape \citep{Bett+07,fw10}, using the halo spin parameter $\lambda$ and minor-to-major axis ratio $c/a$ of the halo shape ellipsoid, which are part of the default \textsc{rockstar} output catalogs. To avoid clutter, we do not display these results. We next deconstruct the assembly bias signal as a function of tidal environment.

\subsection{Assembly bias and tides}
\label{subsec:assemblybias-tides}
\noindent
To begin with, we simply ask what happens to the assembly bias signal when haloes are split by their environment at $4R_{\rm 200b}$. The \emph{left panel} of Figure~\ref{fig:assemblybias-env} is formatted similarly to Figure~\ref{fig:tradnalassemblybias}, except that the larger (blue) symbols joined by thicker lines correspond to filamentary haloes and the smaller (yellow) symbols with thinner lines to node haloes. 
Clearly, filamentary haloes are more strongly clustered -- more biased -- than node halos of the same mass. Although we do not show it here, the dependence of $b_1$ on the tidal environment is much stronger than when, e.g., halo shape is used (c.f. discussion at end of section~\ref{subsec:tradest}). The increase of $b_1$ as the environment becomes anisotropic is in good agreement with previous work \citep[e.g.][]{hahn+09,bprg17}. On the other hand, \citet{fw10} report that $b_1$ decreases as anisotropy increases.  Although their environmental classification is based on the velocity- rather than tidal-shear, so quantitative differences might be expected, the qualitative difference in conclusions is surprising.  We are in the process of checking if they simply mis-stated the correspondence between the velocity-shear based quantities they measured and the sphericity/isotropy of the environment.

\begin{figure}
\centering
\includegraphics[width=0.45\textwidth]{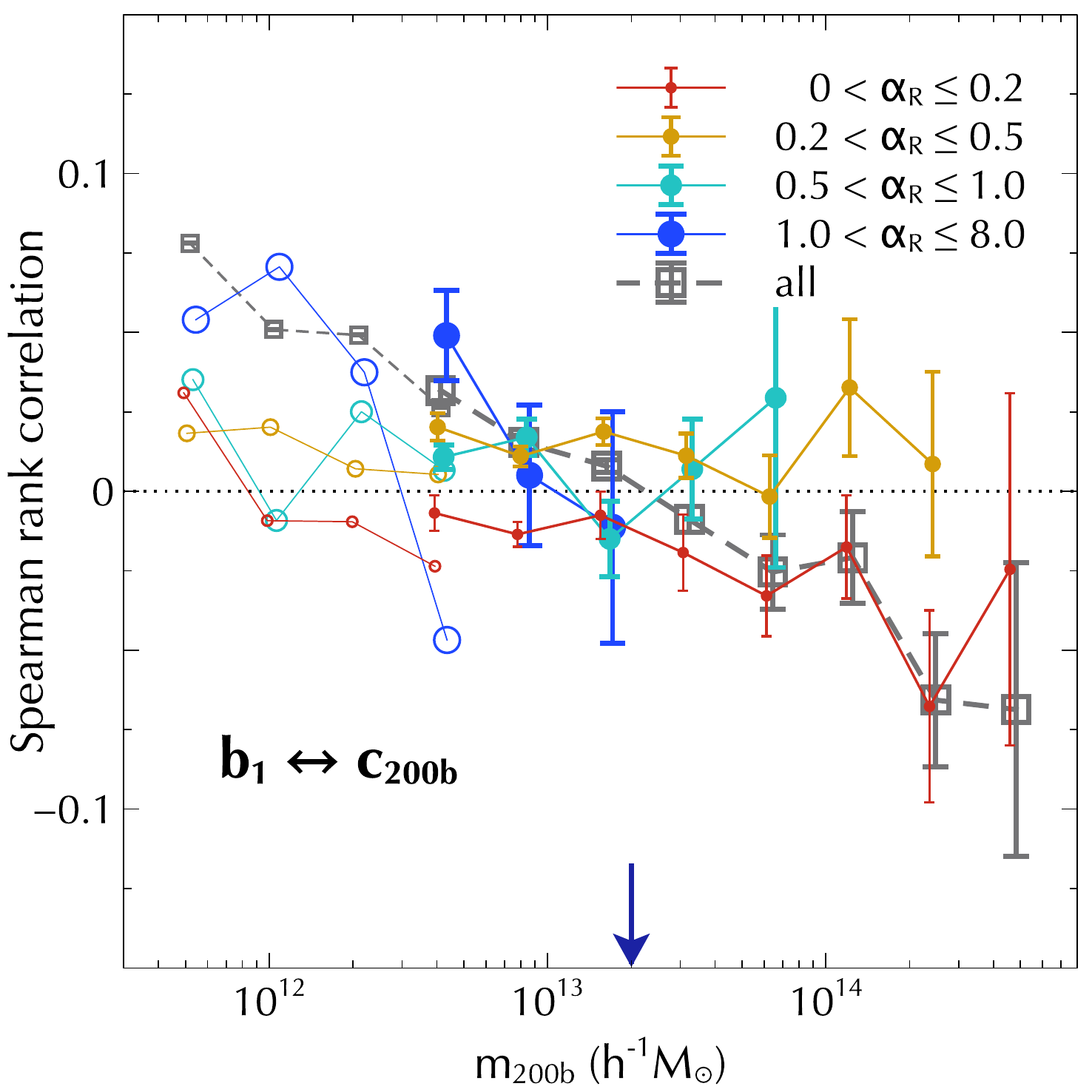}
\caption{Spearman rank correlation coefficient between $b_1$ and concentration $c_{\rm 200b}$, in bins of tidal anisotropy $\alpha_R$ at $R=R_{\rm G,eff}^{(4R_{\rm 200b})}$. Note the difference in vertical scale as compared to Figures~\ref{fig:b1Vsmass-delta} and~\ref{fig:rankcorr-248R200b}. Filled circles of increasing size correspond to increasing values of $\alpha_R$ as indicated, with the colour coding being identical to that in the right panel of Figure~\ref{fig:assemblybias-env}. Additionally, the empty gray squares show the result for all haloes. Results were averaged over \NSIM\ realisations of the default box and the error bars show the corresponding standard error on the mean. The empty symbols joined by thin lines extending to low masses show the results of the single high resolution box. For clarity, measurements in each bin of $\alpha_R$ were given small horizontal offsets. The arrow marks the characteristic mass $m_\ast$ obtained from the peak of the halo mass distribution function. We see that the \emph{sign} of the correlation at fixed $\alpha_R$ is nearly independent of halo mass and goes from negative to positive as $\alpha_R$ increases beyond $\sim0.2$. The overall trend for the all-halo sample at any fixed halo mass thus emerges as an average over halo populations with very different local tidal anisotropy. The criterion $\alpha_R<0.2$ isolates a population of haloes closest to that described by the simplest excursion sets / peaks theory models \citep{dwbs08,ms12}. See text for further discussion.}
\label{fig:rankcorr-b1s-alpha}
\end{figure}

In addition to the strong dependence on the anisotropy of the environment, filamentary haloes show a substantial assembly bias effect at nearly all masses probed, with high concentration haloes being more strongly clustered than low concentration ones (although this trend becomes quite noisy for $m\gtrsim m_\ast$ where the abundance of these haloes is smaller). At the largest masses, the population is dominated by node haloes which, as expected, show the same trend as seen in Figure~\ref{fig:tradnalassemblybias} at high masses. The interesting point to note is that \emph{low mass} node haloes continue to show the same trend as high mass node haloes, with \emph{no inversion} around $m\sim m_\ast$. There is some hint of inversion at the smallest masses, where the signal strength considerably weakens. 

To probe these environmental effects further, in the \emph{right panel} of the Figure, instead of the node/filament split, we use bins of $\alpha_R$ with $R=R_{\rm G,eff}^{(4R_{\rm 200b})}$, as in Figure~\ref{fig:dndlnm-alpha4R200b}. Similarly to the left panel, in each bin we show the mean $b_1$ as a function of mass for all haloes in the bin (circles) and for haloes in the upper and lower quartiles of concentration in that bin (upward and downward triangles, respectively). The all-halo results show a monotonic increase of $b_1$ with $\alpha_R$ at all masses (see also Figure~\ref{fig:b1-mass-alpha4R200b} which showed the median trend; this is also consistent with the positive correlation seen in Figure~\ref{fig:rankcorr-248R200b}). The results split by concentration clearly show that the low mass assembly bias trend is quite sensitive to the value of $\alpha_R$, revealing a rather nuanced set of trends as a function of mass, $\alpha_R$ and concentration. 

The magnitude of the trend between bias and concentration at fixed $\alpha_R$ is quite small for low $\alpha_R$ and becomes noisy for both high $\alpha_R$ and at high masses.
These trends are therefore more easily described using an alternate representation of these results focusing on the \emph{strength} of assembly bias. In Figure~\ref{fig:rankcorr-b1s-alpha} we display the Spearman rank correlation between bias $b_1$ and concentration $c_{\rm 200b}$ as a function of halo mass, for haloes split into the same $\alpha_R$ bins as in Figure~\ref{fig:assemblybias-env}. To orient the discussion, note that, as expected, the all halo result (gray squares) shows a negative correlation at high masses which reverses sign and becomes positive at $m\lesssim m_\ast$ (c.f. Figure~\ref{fig:tradnalassemblybias}). 

We see that there is essentially no \emph{mass} dependence of the $b_1\leftrightarrow c_{\rm 200b}$ correlation for any $\alpha_R$, except at the highest masses for $\alpha_R<0.2$ (where the correlation becomes more negative) and at the lowest masses for $\alpha_R>1.0$ (where the correlation becomes more positive). The \emph{sign} of the signal, however, goes from negative to positive as $\alpha_R$ increases beyond $\sim0.2$ at $m<m_\ast$, while at higher masses the signal becomes consistent with zero for $\alpha_R>0.2$. The trend seen in node haloes in the left panel of Figure~\ref{fig:assemblybias-env} is therefore revealed to be largely driven by haloes in only the most isotropic environments. 
While all the correlations discussed above are quite weak (correlation coefficients $\lesssim0.1$ in magnitude), the correlations are nevertheless statistically significant over a reasonably wide range of $\alpha_R$ and halo mass (e.g., see the error bars for the measurements in the default box for $\alpha_R\lesssim0.5$). At low masses, we see that the all-halo correlation in the high resolution box continues the trend seen in the default box and largely follows that of haloes with $\alpha_R\gtrsim1.0$, while at the highest masses the all-halo correlation follows that of haloes with $\alpha_R\lesssim0.2$.

We also note in passing that, whereas plots such as those in Figure~\ref{fig:assemblybias-env} can be made using traditional estimators of halo bias, the rank correlation measurements in Figure~\ref{fig:rankcorr-b1s-alpha} (and Figure~\ref{fig:rankcorr-248R200b}) are only possible with a halo-by-halo estimator of bias such as $b_1$. Figures~\ref{fig:assemblybias-env} and~\ref{fig:rankcorr-b1s-alpha} form the main results of this work, which we discuss in section~\ref{sec:discuss} below.

\section{Discussion}
\label{sec:discuss}
\noindent
In this section, we discuss in some detail the implications of the results presented in this work.

\subsection{Role of tidal anisotropy in determining assembly bias}
\label{subsec:tidalassemblybias}
\noindent
The main idea we have explored in this work is that a halo's tidal environment is expected to play a significant role in determining its mass assembly history. Our definition of tidal anisotropy $\alpha_R$ (equation~\ref{eq:alpha-def}) evaluated at the Gaussian equivalent of $4R_{\rm 200b}$ (i.e., in the local halo environment) allows us to statistically quantify this connection, as we discuss next.

We have seen (Figure~\ref{fig:rankcorr-248R200b}) that $\alpha_R$ is a better indicator of the \emph{large scale} environment of haloes at fixed mass than is the density contrast $\delta_R$ smoothed on the same scale $R\sim4R_{\rm 200b}$. Specifically, haloes that live in anisotropic local environments tend to cluster more strongly than haloes of similar mass in more isotropic local environments (Figure~\ref{fig:b1-mass-alpha4R200b} and right panel of Figure~\ref{fig:assemblybias-env}). The variable $\alpha_R$ also has the nice property that it sharply distinguishes between node and filament environments as defined by counting the number of positive eigenvalues of the tidal tensor smoothed on the same scale, with the segregation occurring at $\alpha_R\simeq0.5$ independent of halo mass (Figure~\ref{fig:alpha4R200b-mass}). 

Our main aim has been to understand the inversion of the halo assembly bias trend at low masses, where more concentrated haloes are clustered more strongly than less concentrated ones. 
This is the opposite of the trend predicted by simple peaks theory or excursion set models, which is in fact qualitatively realised at \emph{high} masses. Previous studies suggest that this inversion is likely to be associated with the varying tidal environments of low mass haloes \citep{hahn+09}; the strong tidal forces in filamentary environments can quench halo growth, resulting in old, small haloes in highly clustered regions \citep{bprg17}. We have seen in the left panel of Figure~\ref{fig:assemblybias-env}, in fact, that the low-mass inverted trend is largely restricted to haloes classified as being in filaments at $4R_{\rm 200b}$. 

The tidal anisotropy $\alpha_R$ turns out to be a better indicator of the strength of the assembly bias signal than the node/filament split, functioning like a continuous knob rather than a binary switch that controls the sign and strength of the signal (right panel of Figure~\ref{fig:assemblybias-env}, and Figure~\ref{fig:rankcorr-b1s-alpha}). We see in Figure~\ref{fig:rankcorr-b1s-alpha} that, for low mass haloes in the most isotropic environments ($\alpha_R\lesssim0.2$), halo bias and concentration are weakly but significantly \emph{anti}-correlated, just like for their high mass counterparts. The criterion $\alpha_R<0.2$ has therefore isolated a population of haloes that is perhaps closest to that described by the simplest excursion sets / peaks theory models which ignore environmental anisotropy. It will be interesting to check whether the mass function of such objects is more universal than that of the full population of haloes \citep{Tinker08}. The high mass end of this population contains the usual massive cluster-sized haloes, whereas the low mass end is dominated by objects in large scale sheets and voids (Figure~\ref{fig:isowebfracn}; also c.f. the visualisation in Figure~\ref{fig:halovisual}). 

As the tidal anisotropy increases beyond $\alpha_R\simeq0.2$, this small negative correlation turns positive (or, within the noise, consistent with zero at higher masses.) Overall, if we ignore any correlation between $\alpha_R$ and halo concentration, the traditional low-mass positive assembly bias signature in Figure~\ref{fig:tradnalassemblybias}, as well as the all-halo result in Figure~\ref{fig:rankcorr-b1s-alpha}, can be understood as arising because (a) the signal strength depends on $\alpha_R$ and (b) the fraction of low-mass haloes in environments with a negative signal ($\alpha_R\lesssim0.2$) is subdominant (Figure~\ref{fig:alpha4R200b-mass}). 

The fact that the halo mass dependence of the assembly bias correlation strength largely disappears at fixed $\alpha_R$ in Figure~\ref{fig:rankcorr-b1s-alpha} emphasizes that the tidal anisotropy plays a key role in determining the nature of assembly bias. This dependence of assembly bias sign and strength on tidal anisotropy strongly supports the idea that local tides dominate the mass assembly history of low mass haloes. Low mass haloes are comprised of two populations -- those in highly isotropic environments which behave like `standard' peaks-theory/excursion set haloes and those in anisotropic environments which show a \emph{positive} correlation between concentration (or age) and large scale density.

\subsection{Comparison with previous work}
\label{subsec:compareyang}
\noindent
Recently, \citet{yang+17} have also explored halo clustering and assembly bias as a function of web environment, using a traditional web classification based on the number of positive eigenvalues of the density Hessian $\p_i\p_j\delta_{2\Mpch}$, rather than the tidal tensor. Similarly to the present work, they find that haloes in what they classify as filamentary environments cluster more strongly than average. However, a more detailed comparison with their work reveals several differences too.

E.g., \citet{yang+17} find that haloes classified as being in sheets and voids by their analysis have cross-correlation based bias values that are much larger than those in filaments and nodes (their Figure 8). This is very different from our results which suggest that haloes in the most isotropic environments ($\alpha_R\lesssim0.2$) should have smaller (even negative) bias values than those in anisotropic ones (Figure~\ref{fig:b1-mass-alpha4R200b} and right panel of Figure~\ref{fig:assemblybias-env}). Similarly, while the assembly bias trends reported by \citet{yang+17} for their filamentary haloes are qualitatively consistent with those for our anisotropic environments, they also find a strong mass-dependent assembly bias trend for their node-haloes that is similar to their filamentary haloes (their Figure 13); this has no obvious counterpart in our analysis which finds relatively uniform and substantially weaker assembly bias trends with halo mass at fixed low values of $\alpha_R$ (Figure~\ref{fig:rankcorr-b1s-alpha}).

We believe most of these differences can be attributed to our different classification choices that lead to different halo populations being labelled, e.g., as filaments or nodes. As we have argued above, we believe our segregation based on tidal anisotropy $\alpha_R$ is a particularly useful way of understanding assembly bias trends, and also agrees with other analyses that used the tidal tensor for web classification \citep{hahn+09,bprg17}.

\subsection{Consequences for analytical models}
\label{subsec:analytical}
\noindent
There has been some analytical work on modelling the role of tidal effects on halo abundances and clustering \citep{sams06,scs13,cphs17}.  These studies fall within the context of the excursion set approach, so they attempt to model how tidal effects in the \emph{initial} field affect halo formation.  \citet{sams06} focused on the roles played by the initial ellipticity and prolateness -- in addition to the initial overdensity -- of the protohalo patches which are destined to form virialized halos, whereas the more recent work has studied the role played by the initial tidal shear $q$ associated with the protohalos.  

Our work suggests two important modifications to such studies:  one is that $\alpha$ in the \emph{evolved} field is the more relevant variable, and the other is that the relevant scale for these tidal effects may be \emph{larger} than that of the protohalo.  It will be interesting to see if, with these modifications, such excursion set based studies are able to exhibit the strong trends with $\alpha$ that are apparent in Figures~\ref{fig:dndlnm-alpha4R200b} and~\ref{fig:assemblybias-env}.  Moreover, although these previous studies have considered how bias depends on, e.g., $q$, they have not studied the correlation between the initial tidal shear of the protohalo patch and the concentration of the final halo -- i.e., the  additional assembly bias effect we highlighted in Figure~\ref{fig:rankcorr-b1s-alpha}.  In a forthcoming paper (Musso et al., in preparation), we demonstrate how these populations can be analytically described in modified excursion set models.

\subsection{Implications for observational samples}
\label{subsec:obsvns}
\noindent
We conclude this section with a brief discussion of potential applications of our analysis to real data. On the observational front, there has been considerable recent work on estimating the velocity and tidal fields in our local volume using, e.g., the Sloan Digital Sky Survey (SDSS) \citep{wmyv12,jw13,libeskind+15,hptc17,phct17} and performing constrained simulations of the local volume \citep{sorce+16,wang+16}. In the context of our analysis above, the galaxy group-based algorithm of \citet{wmyv12} is of particular interest, since the tidal field information derived from this algorithm could be used to calculate $\alpha_R$ for individual SDSS galaxy groups. 

Given the connection between $\alpha_R$ and assembly bias that we have established in this work, we expect such an analysis to provide us an interesting new handle on \emph{galaxy} assembly bias. Recent analyses of the SDSS main sample have shown that the observed level of assembly bias -- as quantified by the correlation between large scale density and the fraction of (central) galaxies at fixed luminosity or stellar mass that are quiescent -- is substantially below what is expected from the simplest models connecting galaxies to dark matter haloes \citep[see, e.g.][]{twcm17}. While this could be due to underestimated scatter in the galaxy-dark matter connection in these models \citep[see, e.g.,][for high-resolution hydrodynamical simulations of an albeit small sample of galaxies]{rgbp17}, it might also be the case that the SDSS sample does not sufficiently probe the anisotropic environments that dominate the theoretical signal. An analysis that accounts for local tidal anisotropy could conceivably distinguish between these possibilities. Another application of the tidal anisotropy could be to select environments sampling a broad range of large scale bias (c.f. Figure~\ref{fig:assemblybias-env}), which is relevant for multi-tracer analyses that aim to constrain primordial non-Gaussianity and/or detect large scale relativistic effects \citep[see, e.g.][]{mCds09,hsd11,fcsm15}. We intend to address these issues in the near future.

\section{Conclusions}
\label{sec:conclude}
\noindent
We have explored in detail the correlations between halo properties (mass and concentration) and halo environment, both local and large scale. In particular, we have quantified the nature of halo assembly in different environments by dissecting this signal according to the \emph{anisotropy} $\alpha_R$ of the local tidal environment. Employing a novel halo-by-halo estimator of large scale bias, we have explored the correlations between halo properties and large scale bias, as a function of this local tidal anisotropy.

The picture that emerges from our multi-scale analysis involves \emph{low mass} haloes varying between two regimes of local tidal anisotropy $\alpha_R$. At one end are haloes in highly isotropic local environments ($\alpha_R\lesssim0.2$), corresponding to underdensities at larger scales. These behave like scaled-down versions of their high mass counterparts, dominating their immediate environments and showing age-environment correlations qualitatively consistent with simple spherically averaged analytical expectations (namely, a negative correlation between concentration/age and large scale density). On the other side  ($\alpha_R\gtrsim0.2$) are haloes that live close to and are dominated by more massive objects, progressively more so with increasing $\alpha_R$. These small haloes have highly \emph{anisotropic}, filament-like local environments and show a \emph{positive} correlation between concentration and large scale density. 

The transition between isotropic and anisotropic environments, from the point of view of assembly bias strength, occurs at $\alpha_R\simeq0.2$, which is below the threshold $\alpha_R\simeq0.5$ demarcating the split between the more traditional definition of nodes and filaments (Figure~\ref{fig:alpha4R200b-mass}). Figure~\ref{fig:halovisual} can be reviewed in this new light, with `anisotropic' environments for the low mass haloes now corresponding to $\alpha_R\gtrsim0.2$ (orange to blue circles). While we have focused on parent haloes in this work, it will be interesting to probe the behaviour of halo substructure as a function of $\alpha_R$; in particular, whether $\alpha_R$ could be used as a discriminator of the population of so-called `backsplash' haloes \citep{gkg05}, which ought to have the highest values of $\alpha_R$. We will explore this in future work, along with an extension of our analysis to higher redshifts.

\section*{Acknowledgements}
AP gratefully acknowledges use of computing facilities at IUCAA, Pune. The research of AP is supported by the Associateship Scheme of ICTP, Trieste and the Ramanujan Fellowship awarded by the Department of Science and Technology, Government of India. OH acknowledges funding from the European Research Council (ERC) under the European Union's Horizon 2020 research and innovation programme (grant agreement No. 679145, project `COSMO-SIMS'). AP thanks OCA, Nice for hospitality while part of this work was completed. We thank Marcello Musso for useful discussions and an anonymous referee for a constructive and helpful report.

\bibliography{masterRef}

\appendix

\section{Useful scaling relations}
\label{app:scaling}
\noindent
In this Appendix we note down some useful scalings between various quantities that define our $N$-body simulations and their analysis. 

\subsection{Halo mass and grid}
\label{app:mass-grid}
\noindent
Consider a simulation in a cubic box with comoving length $L_{\rm box}$, matter density parameter $\Omega_{\rm m}$ and number of particles $N_{\rm p}$. Since the critical density at present epoch is $\rho_{\rm crit,0}=3H_0^2/(8\pi G)=2.7754\times10^{11}\Mh/(\Mpch)^3$, a halo resolved with $N_{\rm p}^{\rm (halo)}$ particles will have a mass $M_{\rm halo}$ given by
\begin{align}
M_{\rm halo} &= 3.8524\times10^{11}\Mh \,\left(\frac{N_{\rm p}^{\rm (halo)}}{200}\right) \,\left(\frac{1024^3}{N_{\rm p}}\right) \notag\\
&\ph{3.8524\times10^{11}}\times
\left(\frac{\Omega_{\rm m}}{0.276}\right) \,\left(\frac{L_{\rm box}}{300\Mpch}\right)^3 \,.
\label{eq:Mhalo-Np}
\end{align}
If this mass corresponds to $m_{\rm 200b}$, the mass enclosed in a radius $R_{\rm 200b}$ where the enclosed density is $200$ times the background density, then we can write
\begin{align}
R_{\rm 200b} 
&= 181.7\,\kpch \,\left(\frac{L_{\rm box}}{300\Mpch}\right) \notag\\
&\ph{181.7\,\kpch}\times
\left(\frac{N_{\rm p}^{\rm (halo)}}{200}\right)^{1/3} \left(\frac{1024}{N_{\rm p}^{1/3}}\right)
\label{eq:R200b-Nphalo}\\
&= 678.0\,\kpch \,\left(\frac{M_{\rm halo}}{2\times10^{13}\Mh}\right)^{1/3} \,\left(\frac{0.276}{\Omega_{\rm m}}\right)^{1/3}
\label{eq:R200b-Mhalo}
\end{align}
where we have normalised the halo mass by the $z=0$ characteristic mass for our fiducial cosmology.
If we impose a cubic grid on the box with $N_{\rm g}$ cells for post-processing, then the comoving length $\Delta x = L_{\rm box}/N_{\rm g}^{1/3}$ of each grid cell can be written as
\be
\Delta x = 585.9\,\kpch \,\left(\frac{L_{\rm box}}{300\Mpch}\right) \,\left(\frac{512}{N_{\rm g}^{1/3}}\right)\,.
\label{eq:Deltax}
\ee
The number of these grid cells enclosed in a sphere of radius $2R_{\rm 200b}$ centered on a halo is given by
\begin{align}
N_{\rm encl}(<2R_{\rm 200b}) &\equiv (4\pi/3)(2R_{\rm 200b})^3/(\Delta x)^3\notag\\ 
&= 1\times \left(\frac{N_{\rm p}^{\rm (halo)}}{200}\right) \,\left(\frac{1024^3}{N_{\rm p}}\right) \,\left(\frac{N_{\rm g}}{512^3}\right)\,.
\label{eq:Nencl2R200b}
\end{align}
We can then write
\begin{align}
M_{\rm halo} &= 3.0819\times10^{12}\Mh \,\left(\frac{N_{\rm encl}(<2R_{\rm 200b})}{8}\right) \notag\\
&\ph{10^{12}}\times
\left(\frac{512^3}{N_{\rm g}}\right) \,\left(\frac{\Omega_{\rm m}}{0.276}\right) \,\left(\frac{L_{\rm box}}{300\Mpch}\right)^3\,.
\label{eq:Mhalo-Nencl}
\end{align}
For the configuration we use in the main text ($N_{\rm g}=512^3$, $\Omega_{\rm m}=0.276$, $L_{\rm box}=300\Mpch$), demanding that twice $R_{\rm 200b}$ for a halo be resolved with at least $8$ grid cells then gives a minimum halo mass of $m_{\rm min}\simeq3.1\times10^{12}\Mh$ or $N_{\rm p}^{\rm (halo)}\geq1600$. Note that this grid is used only in post-processing the simulation, and is much coarser than the $2048^3$ grid used for PM calculations in the simulation.

\subsection{Gaussian smoothing}
\label{app:gauss-smooth}
\noindent
In practice, we will use Gaussian smoothing kernels to define, e.g., the tidal tensor in the simulation. Since the widths of Gaussian and Tophat smoothing windows are different for the same smoothing radius, one must be careful to account for this difference. This is most easily done by Taylor expanding the Fourier transform of each window and matching the first non-trivial term in each (proportional to $k^2R^2$). This gives the relation 
\be
R_{\rm G}\approx R_{\rm TH}/\sqrt{5}
\label{eq:RG-RTH}
\ee
Denoting the Gaussian equivalent of $2R_{\rm 200b}$ by $R_{\rm G,eff}^{(2R_{\rm 200b})}=2R_{\rm 200b}/\sqrt{5}$, we find the relations
\begin{align}
R_{\rm G,eff}^{(2R_{\rm 200b})} 
&= 162.5\,\kpch \,\left(\frac{L_{\rm box}}{300\Mpch}\right) \notag\\
&\ph{162.5\,\kpch}\times
\left(\frac{N_{\rm p}^{\rm (halo)}}{200}\right)^{1/3} \,\left(\frac{1024}{N_{\rm p}^{1/3}}\right)
\label{eq:RGeff-Nphalo}\\
&= 606\,\kpch  \,\left(\frac{M_{\rm halo}}{2\times10^{13}\Mh}\right)^{1/3} \,\left(\frac{0.276}{\Omega_{\rm m}}\right)^{1/3}
\label{eq:RGeff-Mhalo}\\
&= 325\,\kpch \,\left(\frac{N_{\rm encl}(<2R_{\rm 200b})}{8}\right)^{1/3} \notag\\
&\ph{325\,\kpch}\times
\left(\frac{512}{N_{\rm g}^{1/3}}\right) \,\left(\frac{L_{\rm box}}{300\Mpch}\right)\,.
\label{eq:RGeff-Nenc}
\end{align}
Note that, for any constant $K$, we have 
\be
R_{\rm G,eff}^{(KR_{\rm 200b})} = (K/2)R_{\rm G,eff}^{(2R_{\rm 200b})}\,.
\label{eq:RGeff-K}
\ee
In the main text, we require the tidal environment at the Gaussian equivalent of $2R_{\rm 200b}$, $4R_{\rm 200b}$, etc. In practice, these are calculated by first measuring the tidal tensor using a series of fixed Gaussian radii and then interpolating the results to the scale corresponding to each halo using, e.g., \eqn{eq:RGeff-Mhalo}.  Equation~\eqref{eq:RGeff-Nenc} then says that the minimum Gaussian radius in this series of windows should be $325\kpch$ when considering haloes resolved by at least $8$ grid cells inside $2R_{\rm 200b}$. We have checked that the results have safely converged when using $15$ equi-log spaced values of Gaussian radius for the interpolation.

\section{Analytical arguments}
\label{app:analytical}
\noindent
In this Appendix, we present some simple analytical arguments that clarify some of the trends discussed in the main text.

\subsection{Correlation between $b_1$ and large scale density}
\label{subapp:b1-delta}
\noindent
Our definition of halo-by-halo bias $b_1$ in \eqn{eq:b1HbyH} is closely linked to the density field filtered on large scales using a sharp filter in Fourier space. This is easily seen by noting that, had we replaced $N_k$ with $P_{\rm mm}(k)$ in that equation -- i.e., weighted by the power spectrum rather than number of $k$-modes -- we would have obtained the ratio of the sharp-$k$ filtered conditional density contrast and the variance of its unconditional counterpart. This also shows that $b_1$ is conceptually identical to the quantity written down by \citet{ps12a} in their equation (22) in the context of deriving halo bias from excursion set models, except that \citet{ps12a} were working with the Lagrangian field while $b_1$ is defined for the Eulerian field \citep[see also][for generalisations to include multiple constraints on the haloes]{cps17,css17}.

We can take this comparison further and understand the size of the scatter in Figure~\ref{fig:b1Vsmass-basic} and the correlation trends in Figure~\ref{fig:b1Vsmass-delta} using some simplified models. To explain the size of the scatter, consider the Gaussian approximation in which the conditional distribution $p(\delta_R|{\rm halo})$ is a Gaussian with mean $\avg{\delta_R|{\rm halo}} = (S_\times/S_{\rm halo})\delta_{\rm halo} = S_\times b_1$ and variance ${\rm Var}(\delta_R|{\rm halo}) = S_R - S_\times^2/S_{\rm halo}$. Here $S_R=\avg{\delta_R^2}$ is the unconditional variance on scale $R\sim1/k_{\rm max}$ for sharp-$k$ filtering, $S_{\rm halo}=\avg{\delta_{\rm halo}^2}$ the unconditional variance on scale $R_{\rm halo}\ll R$ and $S_\times=\avg{\delta_R\delta_{\rm halo}}$ is the cross-correlation term. For sharp-$k$ filtering, we have $S_\times=S_R$, so that $\avg{\delta_R|\delta_{\rm halo}} = S_R b_1$ and ${\rm Var}(\delta_R|\delta_{\rm halo}) = S_R(1-S_R/S_{\rm halo})\simeq S_R$. Since we display $b_1$ in this analogy rather than $S_R b_1$, the variance around the mean should be $S_R(1-S_R/S_{\rm halo})/S_R^2=1/S_R-1/S_{\rm halo}\simeq1/S_R$, with little dependence on halo mass. Setting $k_{\rm max}=0.09h{\rm Mpc}^{-1}$ and using the linear theory power spectrum for our cosmology gives us $S_R\sim0.13$, or a standard deviation of $\sim2.8$ in $b_1$. This compares quite well with the measured standard deviation of $\simeq3$ in $b_1$.

To explain the trends in Figure~\ref{fig:b1Vsmass-delta}, we can simply note that, for large enough $R$, $b_1$ and $\delta_R$ are essentially measuring the same variable and must therefore be positively correlated with a strength that increases with $R$. Alternatively, consider a second simplified model which explicitly models the dependence of $b_1$ on the Gaussian filtered $\delta_R$. In this model, the bias is given by $b_1=\delta_{\rm Lag}/S(R_{\rm Lag})$ where $R_{\rm Lag}$ is the scale which contains mass $\bar\rho(2\pi)^{3/2}R^3(1+\delta_R)$, $S(R_{\rm Lag})$ is the sharp-$k$ variance on this scale and $\delta_{\rm Lag}=\delta_{\rm c}\left(1-(1+\delta_R)^{-1/\delta_{\rm c}}\right)$, which follows from an approximation to spherical collapse. At fixed $R$, as $\delta_R$ increases, $\delta_{\rm Lag}$ and $R_{\rm Lag}$ increase, so that $S(R_{\rm Lag})$ decreases and hence $b_1$ increases, which is qualitatively consistent with the left panel of Figure~\ref{fig:b1Vsmass-delta}. To explain the right panel, we must compute the correlation coefficient between $b_1$ and $\delta_R$ in this model. Assuming that both these variables are Gaussian and Taylor expanding all expressions to $4^{\rm th}$ order in $\delta_R$, we find the correlation coefficient $r_{b_1\delta_R}\simeq1-0.1S_R+0.7S_R^2$ independent of halo mass; as expected, the coefficient approaches unity as $R$ increases.

\subsection{Physical significance of $4R_{\rm 200b}$}
\label{subapp:4R200b}
\noindent
Consider a spherically symmetric overdense region obeying spherical collapse \citep{gg72} and forming a virialised structure of mass $M$, radius $R$ and overdensity $\Delta = \rho(<R)/\bar\rho \approx 200$ at $z=0$, where $\rho(<R) = M/(4\pi R^3/3)$ is the density enclosed inside radius $R$ and $\bar\rho$ is the mean density of the Universe. Let us now ask for the radius $R_{\rm ta}$ of the spherical shell around this halo which is currently (i.e., at $z=0$) decoupling from the Hubble flow and turning around. According to the spherical model, $R_{\rm ta}$ encloses a density $\rho(<R_{\rm ta})\simeq5.5\bar\rho$, so that we can write
\begin{align}
5.5 &\simeq \rho(<R_{\rm ta})/\bar\rho = \frac1{\bar\rho}\left(\frac{M + M_{\rm out}}{4\pi R_{\rm ta}^3/3}\right) \notag\\
&= \Delta\left(\frac{R}{R_{\rm ta}}\right)^3 + \frac{M_{\rm out}/\bar\rho}{4\pi R_{\rm ta}^3/3}\,,
\label{eq:Mta}
\end{align}
where we split the mass enclosed in $R_{\rm ta}$ into the mass $M$ in the halo and the mass $M_{\rm out}$ outside it. If the matter surrounding the halo were unclustered, then we would have $M_{\rm out}/\bar\rho = 4\pi(R_{\rm ta}^3-R^3)/3$, leading to
\be
R_{\rm ta}/R \simeq \left(\Delta/(5.5-1)\right)^{1/3} \simeq 3.5\,.
\label{eq:Rta/R}
\ee
Clustering will increase the value of $M_{\rm out}$ and therefore push $R_{\rm ta}$ to somewhat larger values. Numerical evaluations of the spherical model using reasonable initial density profiles lead to values of $R_{\rm ta}/R$ between $\sim4$-$6$. This could plausibly be related to our finding in the main text that the correlation strength between large scale bias and local tidal anisotropy peaks at around $4R_{\rm 200b}$, nearly independently of halo mass.

\section{Tidal tensor and tidal environment}
\label{app:tidal}
\noindent
The tidal tensor at smoothing scale $R$ (we assume Gaussian smoothing throughout), is defined as
\be
T_{ij}(\mathbf{x}) = \p_i\p_j\psi_R(\mathbf{x})
\label{eq:tidaltensor}
\ee
where the normalised, smoothed gravitational potential $\psi_R(\mathbf{x})$ obeys the Poisson equation 
\be
\nabla^2\psi_R(\mathbf{x})=\delta_R(\mathbf{x})\,.
\label{eq:poisson}
\ee
As described in the main text, the smoothed density contrast $\delta_R(\mathbf{x})$ is obtained in Fourier space as $\delta_R(\mathbf{k})=\delta(\mathbf{k})\e{-k^2R^2/2}$, where $\delta(\mathbf{k})$ is the Fourier transform of the CIC interpolated real space quantity $\delta(\mathbf{x})$. In terms of the Fourier variables above, the tidal tensor is 
\be
T_{ij}(\mathbf{x}) = {\rm FT}\left\{(k_ik_j/k^2)\delta(\mathbf{k})\e{-k^2R^2/2}\right\}\,.
\label{eq:tidaltensor-FT}
\ee 
Denoting the eigenvalues of $T_{ij}$ by $\lambda_1\leq\lambda_2\leq\lambda_3$, the tidal classification of the halo environment at scale $R$ can be summarised as \citep{hahn+07}:
\begin{align}
\lambda_1 > 0 &:\textrm{  node}\notag\\
\lambda_1 < 0\,\,\&\,\,\lambda_2 > 0 &:\textrm{  filament}\notag\\
\lambda_2 < 0\,\,\&\,\,\lambda_3 > 0 &:\textrm{  sheet}\notag\\
\lambda_3 < 0 &:\textrm{  void}
\label{eq:tidalclass}
\end{align}

\section{Connection between halo-by-halo bias and gravitational redshifts}
\label{app:gravredshift}
\noindent
We have shown that the concept of halo-by-halo bias $b_1$ is a very useful way of exploring the relations between a halo's large scale environment and other quantities such as its local tidal environment and internal properties. 
As an aside, we note that our definition of $b_1$ has an interesting connection with gravitational redshifts of galaxy samples that are already being explored observationally \citep{whh11,alam+17}. 

The relative gravitational redshift $\Delta z_{\rm g}(r|\Cal{C}_1,\Cal{C}_2)$ between two galaxy samples selected using criteria $\Cal{C}_1$ and $\Cal{C}_2$, respectively, as a function of pair separation $r$, can be shown to be proportional to 
the integral $\int_{\infty}^r\der x\,x\,(\bar\xi_\times(x|\Cal{C}_1)-\bar\xi_\times(x|\Cal{C}_2))$, where $\bar\xi_\times(x|\Cal{C})$ is the volume averaged cross-correlation at separation $x$ of the sample selected by criterion \Cal{C} with the dark matter field \citep[see, e.g., equations 2-4 of][]{croft13}. Ignoring the contribution of the internal halo profiles of the host haloes to these integrals, we find $\Delta z_{\rm g}(|\vec{x}_1-\vec{x}_2|\,|\Cal{C}_1,\Cal{C}_2) \propto \avg{b_1(\vec{x}_1|\Cal{C}_1)-b_1(\vec{x}_2|\Cal{C}_2)}$, where $b_1(\vec{x}|\Cal{C})$ is the bias of a halo at position $\vec{x}$, selected according to criterion \Cal{C}, and the average is over all haloes selected.

This intimate connection between gravitational redshift and halo-by-halo bias is also visually apparent upon comparing Figure~\ref{fig:halovisual-b1} with Figure 2 of \citet{croft13}, where the author coloured the halo markers with an individual measure $z_{\rm g}$ of the gravitational redshift of each halo with respect to the mean Universe. These Figures are strikingly similar, with the same pattern of volume segregation as a function of $b_1$ or $z_{\rm g}$ visible in the respective plot. While the subsequent analysis by \citet{croft13} used $\Delta z_{\rm g}$ for samples selected by halo mass, our discussion shows that it would be equally interesting to explore other observationally interesting selection criteria. (E.g., selecting by $b_1$ itself, were it possible, would lead to a well-defined constant signal with little scatter.) We will explore this in more detail in future work.

\label{lastpage}

\end{document}